%% file: main.tex
\documentclass[sigconf,nonacm,screen,balance=true]{acmart}

\usepackage{tabularx,booktabs} %
\usepackage{array}     %
\usepackage{makecell}  %
\usepackage{subcaption}
\usepackage{xcolor,colortbl,soul}
\usepackage[linesnumbered,ruled,vlined]{algorithm2e}
\usepackage{enumitem}
\usepackage{bm}
\usepackage{hyperref}
\usepackage[capitalise]{cleveref}

\setcitestyle{square}

\definecolor{vermilion}{rgb}{0.89, 0.26, 0.2}

\newcolumntype{C}[1]{>{\centering\arraybackslash}m{#1}}

\SetKwInOut{Parameter}{Parameter}

\begin{document}

\title{Digital Salon: An AI and Physics-Driven Tool for 3D Hair Grooming and Simulation}

\author{Chengan He}
\authornote{Equal contribution.}
\authornote{Previous Adobe Research intern.}
\affiliation{%
  \institution{Yale University}
  \country{}
}

\author{Jorge Alejandro Amador Herrera}
\authornotemark[1]
\authornotemark[2]
\affiliation{%
  \institution{KAUST}
  \country{}
}

\author{Zhixin Shu}
\affiliation{%
  \institution{Adobe Research}
  \country{}
}

\author{Xin Sun}
\affiliation{%
  \institution{Adobe Research}
  \country{}
}

\author{Yao Feng}
\affiliation{%
  \institution{Max Planck Institute for Intelligent Systems}
  \country{}
}
\affiliation{
  \institution{ETH Z\"{u}rich}
  \country{}
}

\author{S\"{o}ren Pirk}
\affiliation{%
  \institution{Kiel University}
  \country{}
}

\author{Dominik L. Michels}
\affiliation{%
  \institution{KAUST}
  \country{}
}

\author{Meng Zhang}
\affiliation{%
  \institution{Nanjing University of Science and Technology}
  \country{}
}

\author{Tuanfeng Y. Wang}
\affiliation{%
  \institution{Adobe Research}
  \country{}
}

\author{Julie Dorsey}
\affiliation{%
  \institution{Yale University}
  \country{}
}

\author{Holly Rushmeier}
\affiliation{%
  \institution{Yale University}
  \country{}
}

\author{Yi Zhou}
\authornote{Corresponding author.}
\affiliation{%
  \institution{Adobe Research}
  \country{}
}

\renewcommand{\shortauthors}{C. He et al.}

\begin{abstract}
We introduce Digital Salon, a comprehensive hair authoring system that supports real-time 3D hair generation, simulation, and rendering. Unlike existing methods that focus on isolated parts of 3D hair modeling and involve a heavy computation process or network training, Digital Salon offers a holistic and interactive system that lowers the technical barriers of 3D hair modeling through natural language-based interaction. The system guides users through four key stages: text-guided hair retrieval, real-time hair simulation, interactive hair refinement, and hair-conditioned image generation. This cohesive workflow makes advanced hair design accessible to users of varying skill levels and dramatically streamlines the creative process in digital media with an intuitive, versatile, and efficient solution for hair modeling. User studies show that our system can outperform traditional hair modeling workflows for rapid prototyping. Furthermore, we provide insights into the benefits of our system with future potential of deploying our system in real salon environments.
More details can be found on our project page: \url{https://digital-salon.github.io/}.
\end{abstract}


\input{fig/latex/teaser}

\maketitle

\input{sec/1_intro}
\input{sec/2_related}

\input{sec/3_method}

\input{sec/4_exp}
\input{sec/5_conclusion}

\bibliographystyle{ACM-Reference-Format}
\bibliography{main}

\appendix

\end{document}

%% file: fig/latex/teaser.tex
\begin{teaserfigure}
    \centering
    \includegraphics[width=\linewidth]{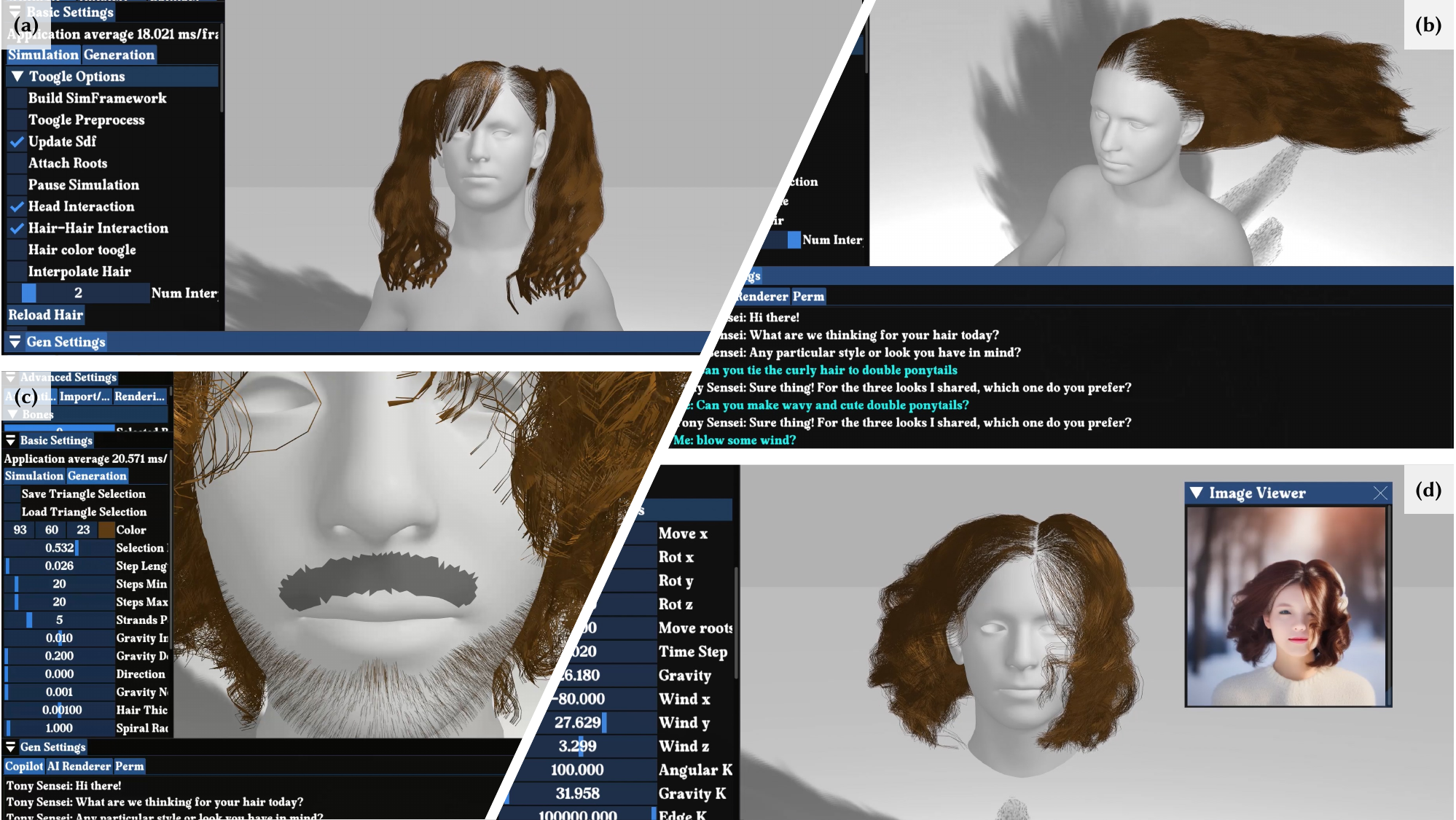}
    \caption{The Digital Salon system. Digital Salon enables users to: (a) retrieve 3D hair models from a curated database using natural language input, (b) simulate hair dynamics in real time, (c) refine fine-grained details through interactive grooming tools, and (d) generate photorealistic images conditioned on the current hairstyle and additional textual descriptions.}
    \label{fig:teaser}
\end{teaserfigure}

%% file: sec/1_intro.tex
\section{Introduction}

Have you ever walked out of a hair salon with a hairstyle that did not match your expectations? Have you ever found that your verbal or visual description led to a hairstyle completely different from what you had imagined? This all-too-common experience underscores a fundamental challenge in hairstyling: how to bridge the gap between a client's description and the mental image formed by the stylist. Although this process is often trial-and-error, progress in 3D hair modeling technologies raises a compelling question: \emph{can we streamline this experience with an interactive tool that enables rapid prototyping of the 3D hairstyles imagined by clients?}

However, in the current production workflow, 3D hair models are still predominantly created manually using advanced software such as XGen, Ornatrix, and Hairfarm. Although these tools offer state-of-the-art features for designing intricate hairstyles, they are tailored for highly skilled digital artists and come with a steep learning curve. Even for experienced professionals, the hair modeling process is highly labor-intensive: it typically requires $6$ -- $10$ working hours to produce low-poly hairstyles represented with hair strips, $16$ -- $24$ hours for stylized hairstyles with individually modeled strands, and $40$ -- $56$ hours for AAA-quality character hairstyles. These statistics underscore the significant time and effort required in traditional hair modeling pipelines.

Recent research has introduced several advancements to lower the technical barriers of 3D hair modeling, including simplifying hair generation by training generative models on portrait images~\cite{zhang2019hair} and text prompts~\cite{Sklyarova_2024_CVPR}, improving simulation efficiency through physics-based~\cite{daviet2023interactive,hsu2024hairinterp} and deep learning-based~\cite{zhou2023groomgen,lyu2020real} frameworks, and enhancing hair rendering quality using neural representations such as implicit volumes~\cite{Wang_2022_CVPR} and 3D Gaussian splats~\cite{luo2024gaussianhair}.
However, most of these works focus on specific aspects of hair modeling and lack a comprehensive hair authoring system that allows users to rapidly prototype 3D hairstyles and explore creative ideas. Moreover, many of these methods rely on computationally expensive neural network training or physics-based simulations, highlighting the need for an interactive and intuitive system where creators can manipulate and refine hairstyles in real time.

To address these limitations, we propose \emph{Digital Salon}, a comprehensive hair authoring system that integrates cutting-edge deep learning and physics-based algorithms to enable real-time generation, simulation, and rendering of highly detailed hairstyles.
At the core of Digital Salon is our copilot system, ``Tony Sensei'', which allows users to effortlessly create intricate hairstyles, generate realistic hair dynamics, and render vivid images using natural language prompts.
Similar to~\cite{hu2015single}, our system is built upon a curated database of high-quality 3D hairstyles. By captioning these hairstyles with detailed text descriptions, Digital Salon enables realistic hair generation by retrieving the database with user-provided text prompts.
Complementing the hair generation system is a real-time dense hair simulator, adapted from the Augmented Mass-Spring Model (AMS)~\cite{alejandro2024ams}. Although previous methods, such as~\cite{daviet2023interactive}, have made notable progress in dense hair simulation, they are limited to simulating straight hair under simple dynamic conditions. In contrast, the AMS simulator integrated in Digital Salon supports real-time simulation of more individual hair strands, accommodating a variety of hair types from straight to curly.
Additionally, interactive editing algorithms are designed to allow users to refine and customize generated hairstyles in real time.
Finally, Digital Salon incorporates a hair rendering system based on ControlNet~\cite{zhang2023adding}, allowing users to specify additional elements, such as clothing and background, through text prompts. With 3D hairstyles as the structural condition, the rendered images ensure an alignment of the given hairstyle with a cohesive and visually stunning image quality.

To evaluate Digital Salon, we conducted user studies and interviews in which participants were asked to create various 3D hairstyles using our system. Our participants included both digital content creators and a professional hairstylist, and the results indicate that Digital Salon provides intuitive and efficient interactions, enabling users to rapidly explore and iterate on their creative ideas in a digital environment. Moreover, Digital Salon demonstrates strong potential for deployment in real salon settings, offering a more expressive and informative alternative to traditional 2D reference images by incorporating 3D hairstyle visualization, physics-based simulation of hair textures, and interactive feedback.

Our contributions are summarized as follows:
\begin{itemize}
    \item We propose Digital Salon, a comprehensive hair authoring system that seamlessly integrates text-to-hair generation, physics-based simulation, and AI-driven rendering.
    \item We design our system to run in real time on consumer-grade hardware, enabling users to rapidly prototype, edit, and simulate 3D hairstyles with immediate visual feedback. %
    \item Through user studies and interviews, we evaluate the system’s intuitiveness, efficiency, and semantic precision, and explore its potential applications in both digital content creation and real-world salon environments.
\end{itemize}

%% file: sec/2_related.tex
\section{Related Work}

Hair modeling has been studied in computer graphics for several decades. Following the typical workflow to create 3D hair models, existing techniques can be categorized as generation, simulation, and rendering~\cite{ward2007survey}. These are three key stages covered by our system, and in this section we review related literature for each of them.

\subsection{3D Hair Capture and Generation}

In production pipelines, 3D hair models are still primarily crafted using professional tools such as XGen, Ornatrix, and HairFarm -- a process that remains time-consuming and labor-intensive. Even for experienced artists, modeling complex structures like flyaway strands, tight curls, or coils poses significant challenges. As an alternative, hair capture methods aim to reconstruct real-world hairstyles with strand-level detail. However, state-of-the-art capture techniques often rely on specialized equipment such as multi-camera rigs~\cite{paris2004capture,jakob2009capturing,xudynamichair} or even CT scanners~\cite{shen2023CT2Hair}, leaving them inaccessible to most users. To broaden accessibility, deep learning-based methods have emerged with synthetic 3D hair datasets~\cite{hu2015single} to infer 3D hair geometry from single-view~\cite{chai2016autohair,saito2018_hairvae,zhou2018hairnet,zheng2023hair,wu2022neuralhdhair} or sparse-view~\cite{zhang2017data} image input, significantly reducing hardware requirements. 
However, the range of styles produced from these data-driven methods is constrained by the training sets they used, which typically do not include afro-textured, coily, or volumetric hair types.

Due to the absence of large-scale real-world 3D hair datasets, recent hair generative models~\cite{zhou2023groomgen,Sklyarova_2024_CVPR,he2025perm} continue to rely on synthetic data with augmentations. These models often incorporate multimodal conditioning, such as portrait images or text prompts, to enhance generation quality and control. For interactive 3D hair modeling, Xing et al.~\cite{xing2019hairbrush} introduced a VR-based authoring system that enables users to design hairstyles by manually drawing high-level guide strips in 3D space. While pursuing a similar goal of user-driven hair modeling, our system lowers the interaction barrier even further by enabling intuitive control through natural language descriptions.

\subsection{3D Hair Simulation}

Simulating 3D hair dynamics requires accurately modeling hair-hair interactions, elastic deformation, and responses to external forces such as wind and gravity. Physically-based methods have been widely explored to reproduce these complex behaviors, including the Super-Helix model~\cite{bertails2006super}, Discrete Elastic Rods (DER)~\cite{bergou2008discrete}, and formulations using exponential time integrators\cite{michels2015physically,10.1145/3072959.3073706}. 
While these methods produce highly realistic results, their computational cost prohibits their use in real-time applications.
To address these challenges, Daviet et al.~\cite{daviet2023interactive} introduced significant engineering improvements to the DER framework.
Although this advancement marks a substantial step toward interactive hair simulation, DER-based models remain computationally intensive for simulating strand dynamics, making them challenging to apply in more intricate interactive scenarios with complex hairstyles.
The Mass-Spring (MS) model~\cite{rosenblum1991simulating} is another widely used approach for hair simulation, offering a simpler and more computationally efficient alternative to DER-based models. In this method, each strand is represented as a sequence of interconnected masses with elastic constraints. 
However, the standard MS model cannot capture global strand behavior, leading to instability, excessive sagging, and a loss of hair structure during simulation.
To address these limitations, subsequent work introduced modifications such as the incorporation of extra springs to enhance system stability~\cite{alejandro2024ams, selle2008mass} and the development of more efficient solvers and improved hair-body interactions~\cite{jiang2020real}. A notable advancement is the Augmented Mass-Spring Model (AMS) proposed by Amador Herrera et al.~\cite{alejandro2024ams}, which enables real-time simulation of dense hair at the strand level on consumer-grade hardware.

With the rise of machine learning in physics, data-driven approaches have been explored to accelerate hair simulation. One line of research focuses on predicting the shapes of individual hair strands based on factors such as gravity direction and head pose~\cite{zhou2023groomgen}, while another employs deep learning to enhance the fine details of interpolated dense hair from sparse guide strands~\cite{lyu2020real}. Although these methods are effective, their performance is heavily dependent on the distribution of the training data, yielding performance degradation when applied to out-of-distribution dynamic scenarios.

Considering the trade-off between simulation efficiency and fidelity, our system adopts AMS as our hair simulator, achieving real-time hair simulation with decent physical plausibility.

\subsection{3D Hair Rendering}

Realistic 3D hair rendering remains a fundamental challenge in computer graphics due to the complex light transport within dense, thin fiber structures. Traditional techniques rely on physically-based analytic fiber shading models, such as the Kajiya-Kay model~\cite{kajiya1989rendering} and the Marschner model~\cite{marschner2003light}, which consider the variation of reflectance in azimuthal and longitudinal directions and account for internal absorption and caustics. Numerous extensions have been proposed to enhance these models, such as adding a non-physical diffuse term~\cite{zinke2009}, decoupling into artist-friendly lobes~\cite{sadeghi2010}, enforcing energy conservation~\cite{deon2011energy}, adapting to production rendering~\cite{chiang2015practical}, and capturing nuanced phenomena like colorful glints~\cite{xia2023practical}.
While these models are widely used in the game and film industries, their computational complexity hinders deployment in real-time applications.

To improve rendering efficiency, various methods have been proposed to approximate strand-based hair models using simplified representations. These representations include hair mesh textures~\cite{bhokare2024real} and clustered thick strands~\cite{huang2024real}, which reduce computational complexity while preserving visual fidelity with carefully designed shading models and level-of-detail techniques.
Another thread of research leverages neural rendering techniques to enhance the efficiency and photorealism of hair rendering. Neural representations such as implicit volumes~\cite{Wang_2022_CVPR} and 3D Gaussian splats~\cite{luo2024gaussianhair,zakharov2024gh} have been explored to learn complex light transport behavior from multi-view images or monocular videos, enabling high-fidelity hair rendering with reduced computational overhead.

In the context of generative models, diffusion models~\cite{ho2020denoising} have been explored for conditional image synthesis. ControlNet~\cite{zhang2023adding}, built on Stable Diffusion~\cite{rombach2022high}, enables controllable image generation by incorporating structural conditions, making it suitable for hair ``rendering'' conditioned on its geometric structure. Our system builds on this progress by integrating ControlNet-based image generation to allow intuitive text-driven customization, ensuring that generated images are consistent with both the 3D hair structure and the user’s intent.

%% file: sec/3_method.tex
\section{Design Goals}

To inform the design of our system, we conducted an interview-based formative study aimed at understanding users’ past hairstyling experiences and their expectations for digital hair modeling tools. We invited eight creators (C1 -- C8) to participate. Seven of them reported having negative experiences with real-world hairstyling, and their backgrounds in content creation varied significantly. Specifically, two participants (C1 and C6) have novice-level experience with 3D content creation; five participants (C2, C3 -- C5, C7) have intermediate to expert-level experience; and one participant (C8) is a professional stylist with real-world haircutting experience. Additionally, C3 have expertise in hair simulation, and C4 have significant experience in hair rendering.
We conducted these open-ended interviews with two primary goals: (1) to explore participants’ personal hairstyling challenges and frustrations, and (2) to identify the limitations in current digital hair modeling workflows. Participants were asked to describe their past hairstyling experiences, reflect on the limitations of existing tools, and share their vision for an ideal interactive hair modeling system.

The interviews helped us understand the current issues in both real-world styling and digital content creation regarding hair. In real-world scenarios, \textbf{the absence of visual feedback during verbal communication} was unanimously identified as a major factor contributing to unsatisfactory results. Participants noted that clients often lack the vocabulary to describe specific hairstyles (C6), and even descriptive terms are used, their interpretation can vary due to differing aesthetic preferences (C4, C5). Although reference images in magazines offer some visual guidance, they do not capture the back view of the hairstyle due to occlusion, making it difficult to fully convey the intended look (C4).

Another issue raised by the participants was \textbf{the lack of consideration for the physical properties of hair} in the general hairstyling process. As C1 noted, \textit{``You can ask them for the shape you want, but they do not understand the physical texture of how your hair falls, how gravity affects it, and things like that. They cannot make it the same shape if they do not account for the physical property.''} This concern was echoed by C8, a professional stylist, who explained that clients often bring reference images depicting straight hairstyles, even when their natural hair is curly. In such cases, even if the stylist can replicate the style in the salon, it becomes difficult for clients to maintain it on their own. C8 further emphasized that hair texture is not static: it can change with length, humidity, and how the hair is tied. For example, short curly hair tends to become overly voluminous, which some clients may find unappealing. As a result, adapting static hairstyle references to match individual hair textures is both tedious and challenging, and may occasionally lead to client dissatisfaction.

In the context of digital hair modeling, participants primarily expressed concerns related to \textbf{interactivity and efficiency}. C1–C3 said that controlling 3D digital hair is particularly challenging, especially if you want their movement to obey physical laws. Since physics-based simulation is often too slow to support real-time interaction, it is difficult to iteratively refine hairstyles during grooming. Additionally, C4 pointed out that hair rendering remains a computational bottleneck, particularly for light-colored hair, which involves complex light interactions such as multi-bounce scattering. As a result, producing high-quality visual outputs requires significant rendering time, further limiting the efficiency of current digital workflows.

The formative study revealed a strong desire for an interactive hair authoring system that integrates natural language input, 3D visual feedback, and physics-based simulation, all of which can streamline the process of designing, refining, and even physically reproducing hairstyles. Drawing from the insights gathered in these interviews, we distill the following design goals to guide the development of the Digital Salon system:
\begin{enumerate}[label=\textbf{G\arabic*.}]
    \item \textbf{Support natural language input with visual feedback.} The system should allow users to describe hairstyles using natural language while receiving immediate visual feedback in a fully 3D environment, thus offering a more expressive and flexible medium for communicating hair-related ideas compared to 2D reference images.
    \item \textbf{Enable physics-based simulation with customizable parameters.} To preview realistic hair behavior, the system must incorporate physics-based simulation. The physical parameters should be adjustable to accommodate the diverse hair types and behaviors found in real-world scenarios.
    \item \textbf{Provide real-time performance across core functionalities.} The system should maintain real-time responsiveness throughout the key stages of the workflow, including generation, simulation, and rendering, to ensure an interactive authoring experience.
    \item \textbf{Be intuitive and accessible to users with varying technical backgrounds.} Reflecting the diverse expertise observed in our formative study, the system should support both novices and experienced professionals in quickly and effectively creating high-quality 3D hairstyles.
\end{enumerate}

\section{Digital Salon}
\input{fig/latex/workflow}
\subsection{Overview}

We incorporated the findings and design goals from our formative study to guide the development of Digital Salon. As shown in~\cref{fig:interface}, our user interface comprises three main components: (a) the \textbf{3D viewport}, which visualizes the head mesh and 3D hairstyles, (b) the \textbf{parameter panel}, which exposes advanced controls and detailed parameters for experienced users, and (c) the \textbf{interaction panel}, where users can engage with our virtual agent, ``Tony Sensei'', for intuitive, text-based manipulation of 3D hairstyles.
\input{fig/latex/interface}

Our interaction panel features two key components: the \textbf{Copilot tab} and the \textbf{AI Renderer tab}. The Copilot tab provides an interactive window where users can enter text prompts, allowing ``Tony Sensei'' to autonomously perform corresponding operations such as generating 3D hair candidates and simulating hair dynamics under a wind field. The AI Renderer tab enables text-guided rendering, allowing users to specify additional elements like clothing and background. The system then generates an image based on the current 3D viewpoint, seamlessly integrating the presented hair structure with the specified textual descriptions.

\cref{fig:workflow} illustrates the typical workflow users go through to create 3D hairstyles in Digital Salon. Imagine a user, Ayaka, who wants to explore a new hairstyle before an upcoming appointment. She begins by providing a textual description of her desired look, and the virtual agent, ``Tony Sensei'', will quickly generate three 3D hairstyle candidates for her to choose from. Ayaka can rotate and inspect each option from arbitrary viewpoints, with real-time hair dynamics offering a more holistic understanding of the hairstyle with its physical behavior. If adjustments are needed, such as trimming certain sections, she can refine the style interactively using our grooming tools. Once satisfied, Ayaka can invoke the AI renderer to produce a photorealistic image, offering a realistic preview of how the hairstyle would appear in real life. The final 3D hairstyle can then be shared with her stylist for precise reproduction.
Similarly, digital content creators can benefit from the system to rapidly prototype and iterate on 3D hairstyles for various artistic and production needs.
In the following, we delve into the details of each step in the workflow.

\subsection{Text-Guided Hair Retrieval}

Similar to~\cite{hu2015single}, our system is built on a curated database of high-quality 3D hairstyles manually groomed by professional artists. To ensure a balanced representation across various curl patterns, we follow the widely recognized ten hair types\footnote{\url{https://www.healthline.com/health/beauty-skin-care/types-of-hair}} as a guideline during asset creation.
The final database comprises 1,320 hairstyles, spanning a wide range of diverse and complex styles, including dreadlocks, double tails, and double buns. Each hairstyle consists of approximately $10,000$ to $80,000$ strands. To enrich geometric understanding, each hairstyle is rendered from four canonical viewpoints (front, back, left, and right) using Blender, yielding multi-view imagery that captures the full structure of the hair. \cref{fig:database} shows three representative examples, with additional samples available in supplemental.
To facilitate text-to-hair retrieval, these rendered images are concatenated and captioned using InternVL 2.0~\cite{chen2024internvl}, which generates concise and descriptive captions within 60 words. Further details about our image captioning pipeline can be found in supplemental.
\input{fig/latex/database}

For each caption, we use CLIP~\cite{radford2021learning} to tokenize the text and obtain the corresponding token sequence $\bm{t}$, which is then encoded to generate a normalized text embedding $\bm{e}_t \in \mathbb{R}^{512}$:
\begin{equation}
    \label{eq:clip}
    \bm{e}_t = \mathcal{E}_t(\bm{t})\,,
\end{equation}
where $\mathcal{E}_t$ denotes the text encoder in CLIP. The embeddings of all captions in our database are concatenated to form an embedding matrix: $\bm{E}_t = [\bm{e}_t^1, \bm{e}_t^2, \dots, \bm{e}_t^{1320}]^\top \in \mathbb{R}^{1320 \times 512}$, which captures the textual information of all stored hairstyles.

For a user-provided text prompt, it follows the same tokenization and encoding process as in~\cref{eq:clip} to obtain the corresponding embedding $\bm{e}_u$. This embedding is then multiplied by the precomputed embedding matrix $\bm{E}_t$ to compute the cosine similarity between the input query and all captions in the database:
\begin{equation}
    \bm{s} = \bm{E}_t\bm{e}_u^\top \in \mathbb{R}^{1320}\,,
\end{equation}
where $\bm{s}$ is the similarity vector indicating the relevance of each database entry to the user’s input. The system selects the top three highest-scoring entries and retrieves the corresponding hairstyles for user selection. This process is executed in real time, enabling interactive exploration. In~\cref{fig:retrieval}, we present four examples of hairstyles retrieved based on different user inputs.
\input{fig/latex/retrieval}

\subsection{Real-Time Hair Simulation}

To offer a vivid and dynamic preview of the 3D hairstyle’s physical properties, our system integrates real-time hair simulation. This allows users to freely inspect the hairstyle from arbitrary viewpoints and observe its behavior under various dynamic conditions, such as wind and head motion. The simulation is powered by the Augmented Mass-Spring Model (AMS)~\cite{alejandro2024ams}, which supports high-fidelity, strand-level dynamics in real time while remaining efficient enough to run on consumer-grade hardware.

The AMS formulation discretizes the hair assets at the strand level, modeling standard edge, bending, and torsional degrees of freedom while incorporating an augmented biphasic interaction through two additional one-way springs. This extended mass-spring coupling is embedded within a hybrid Eulerian/Lagrangian framework to efficiently resolve hair-hair interactions. Leveraging this method, our system is capable of simulating thousands of interacting strands in real time, across a variety of hairstyles and dynamic conditions. Additionally, our assets are designed to be directly compatible with the AMS framework, enabling on-the-fly discretization and simulation of strands without the need for additional preprocessing.

With this design, users can interactively drag the head mesh to observe how the hair responds to head movements. For advanced users, the parameter panel (see~\cref{fig:interface}) exposes detailed simulation settings, allowing them to adjust parameters and fine-tune the simulation to the hair texture they want. To further streamline interactions, we provide special text-based instructions for ``Tony Sensei''. For example, typing \textit{“create some wind?”} in the interaction panel prompts ``Tony Sensei'' to automatically generate a wind field, allowing users to preview how the hair dynamically reacts to the simulated wind conditions. 
In~\cref{fig:wind}, we illustrate how double tails respond dynamically to a wind field generated from such user input.
Additional demonstrations of our real-time simulation capabilities can be found in the supplementary video.
\input{fig/latex/wind}

\subsection{Interactive Hair Refinement}

Since the retrieved hairstyle may not perfectly match the user’s creative vision, our system allows for interactive refinement through real-time grooming tools. Users can fine-tune the hairstyle using two key operations: \textbf{procedural strand growth}, which enables the addition of new hair features such as beards, mustaches, or other facial and peripheral hair regions; and \textbf{simulation-driven grooming}, which allows users to selectively grab and shorten hair strands based on dynamic behavior or visual preference. These editing tools provide immediate visual feedback, maintaining physical plausibility through integrated simulation, and offer an intuitive and engaging user experience for detailed hairstyle customization.

\subsubsection{Procedural Strand Growth}

Our procedural strand growth algorithm begins with a user-defined set of triangles painted on the head mesh. We sample a root position $\boldsymbol{v}_0$ within the selected triangles and compute the initial strand direction $\boldsymbol{p}^{0}_{\text{dir}}$ by taking a weighted average of the triangle's per-vertex normals using barycentric coordinates, followed by the addition of a small perturbation vector $\boldsymbol{\delta}$ sampled from the uniform distribution $\mathcal{U}(-1,1)$. 

Strand vertices are then generated sequentially, starting from the root. At each step $i$, the strand direction is updated as:
\begin{equation}
    \boldsymbol{p}^{{i}^{'}}_{\text{dir}} = \boldsymbol{p}^{i-1}_{\text{dir}} + \boldsymbol{p}^{i-1}_{\text{grav}}\,\max\left(p_{\Gamma}, 1-\|\boldsymbol{p}^{i-1}_{\text{dir}}\cdot (0,1,0)\|\right)\,,
\end{equation}
where $p_{\Gamma}$ controls the maximum deviation from the vertical axis, and the gravity vector $\boldsymbol{p}^{i}_{\text{grav}}$ introduces downward displacement:
\begin{equation}
    \boldsymbol{p}^{i}_{\text{grav}} = (0,-ip_{\gamma}, 0)\,,
\end{equation}
with $p_{\gamma}$ representing the gravity strength. 

To simulate curl or wavy effects, an additional update step incorporates a helical displacement:
\begin{equation}
    \boldsymbol{p}^{i}_{\text{dir}} = \boldsymbol{p}^{{i}^{'}}_{\text{dir}} + p_{\Omega} \left( \boldsymbol{p}^{{i}^{'}}_{\text{dir}}-\boldsymbol{H}^{i}\right)\,,
\end{equation}
where $p_{\Omega}$ controls the strength of the spiral influence, and the helix vector $\boldsymbol{H}^{i}$ is defined as:
\begin{equation}
    \boldsymbol{H}^{i} = \left(p_{h} \cos{(ip_{\text{freq}})},1, p_{h} \sin{(ip_{\text{freq}})}\right)\,,
\end{equation}
with $p_{h}$ denoting the helix radius and $p_{\text{freq}}$ controlling the curl frequency. A summary of this process is provided in~\cref{algo:growth}. 
\input{algorithm/growth}

In~\cref{fig:growth} we demonstrate the generation capabilities of our procedural growth scheme by performing a parameter space exploration of the helix radius $p_{h}$ and the gravity parameter $p_{\gamma}$, with other parameters set to their default values ($p_\Gamma=0.2$, $p_\Omega=0.3$, $p_{\text{freq}}=1.0$). Furthermore, in~\cref{fig:beard}, we illustrate the application of our technique to facial hair generation, showcasing the creation of diverse styles of beards and mustaches.
\input{fig/latex/growth}

\input{fig/latex/beard}

\subsubsection{Simulation-Driven Grooming}

We further extend the digital grooming capabilities of our framework by introducing a grooming mechanism integrated with our underlying simulation framework. Unlike traditional approaches that rely solely on geometric strand manipulation, our method models user interaction through the application of spring forces, enabling intuitive repositioning of hair into a variety of configurations. Moreover, trimming is enabled by explicitly removing the particles corresponding to the trimmed region, along with the springs connecting them to the rest of the strand. This induces a physically meaningful response in the simulation, as the removal of springs leads to an imbalance of forces that propagates through the system. As a result, the simulation dynamically adapts to both user-driven motion and topological changes in the hair structure. This approach enables real-time updates that faithfully capture the complex deformation behavior observed in real-world grooming scenarios, as illustrated in~\cref{fig:trim}. Please refer to our supplementary video for the entire grooming process.
\input{fig/latex/trim}

\subsection{Hair-Conditioned Image Generation}

Leveraging advanced image generation models, our system facilitates the creation of high-quality images, where the simply shaded head and hairstyle serve as the structural control. Additional text prompts allow users to specify supplementary elements that are not present in the original rendering, such as clothing or background details.

In the AI Renderer tab, users are provided with a guided interface to enter text prompts describing attributes such as gender, hair color, head pose, and other miscellaneous details, including background and clothing. The system captures the current image from the 3D viewport, crops it, and applies the Canny filter~\cite{canny1986computational} to generate an edge map. This edge map serves as the structural condition for ControlNet~\cite{zhang2023adding}, which, together with the user’s text prompt, guides the generation of the final image.

The generated images are photorealistic and closely aligned with the original hairstyle, enabling users to rapidly and reliably prototype their creations. we present four examples of the conditional hairstyles, extracted edge maps, and final generated images.
\input{fig/latex/controlnet}

\subsection{Implementation Details}

Our core system is implemented in C++ with CUDA, encompassing the user interface and the hair simulation and grooming algorithms. The text-guided hair retrieval and hair-conditioned image generation functionalities are developed using Python 3.10, PyTorch 2.4.1, and CUDA Toolkit 12.1. These Python components are integrated into the C++ system via the Python/C API, enabling communication between the user interface and the Python backend for deep learning-related operations. For hair-conditioned image generation, we employ ControlNet based on the Stable Diffusion 1.5 backbone~\cite{rombach2022high}, utilizing the official checkpoint of ControlNet v1.1 Canny as the control mechanism.

%% file: fig/latex/workflow.tex
\begin{figure*}[ht]
    \centering
    \includegraphics[width=\linewidth]{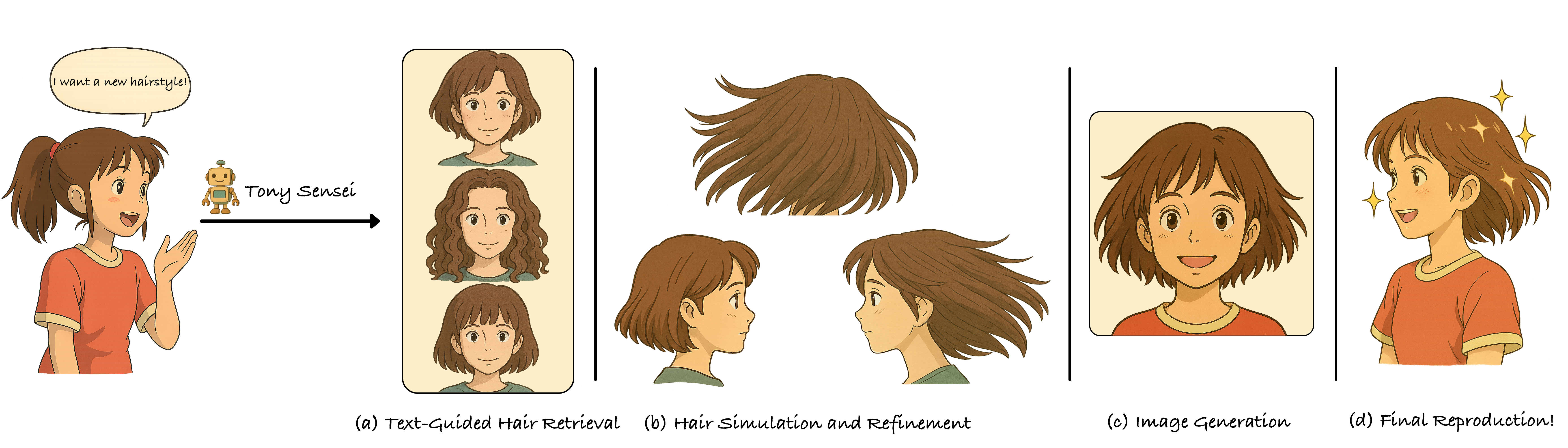}
    \caption{Authoring workflow of Digital Salon. To create a new hairstyle whether for digital content or real-world styling, (a) the user begins by expressing their intent through natural language, prompting the system to retrieve suitable hairstyle candidates from the database. (b) The selected hairstyle can then be interactively simulated and refined to meet specific design requirements. (c) A photorealistic image can be generated to provide a realistic preview of the final look. (d) Once satisfied, the user can share the 3D hairstyle with a stylist for accurate reproduction. The contents in this figure are generated by ChatGPT-4o~\cite{openai2024gpt4o} in Studio Ghibli style.}
    \label{fig:workflow}
\end{figure*}

%% file: fig/latex/interface.tex
\begin{figure}[ht]
    \centering
    \includegraphics[width=\columnwidth]{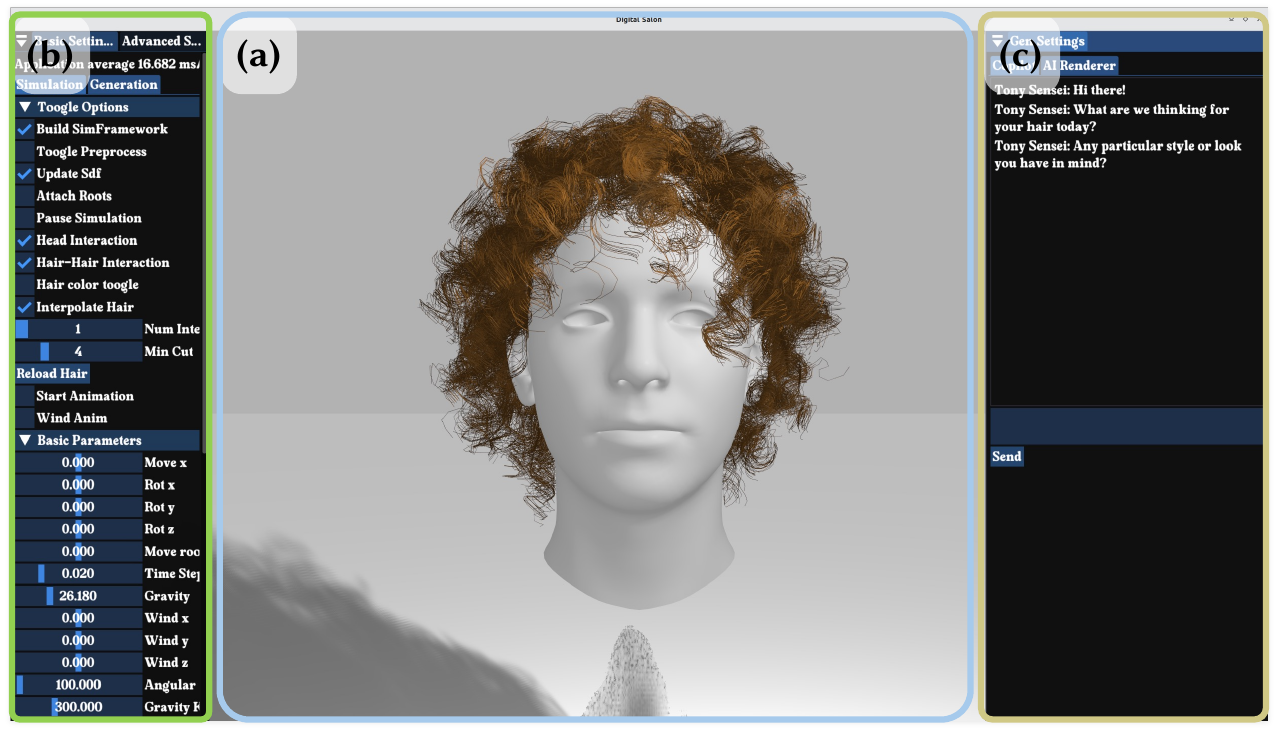}
    \caption{User interface of Digital Salon, comprising: (a) the main 3D viewport for visualizing hairstyles, (b) the parameter panel for detailed settings, and (c) the interaction panel for user input and controls.}
    \label{fig:interface}
\end{figure}

%% file: fig/latex/database.tex
\begin{figure}[ht]
    \centering
    \includegraphics[width=\columnwidth]{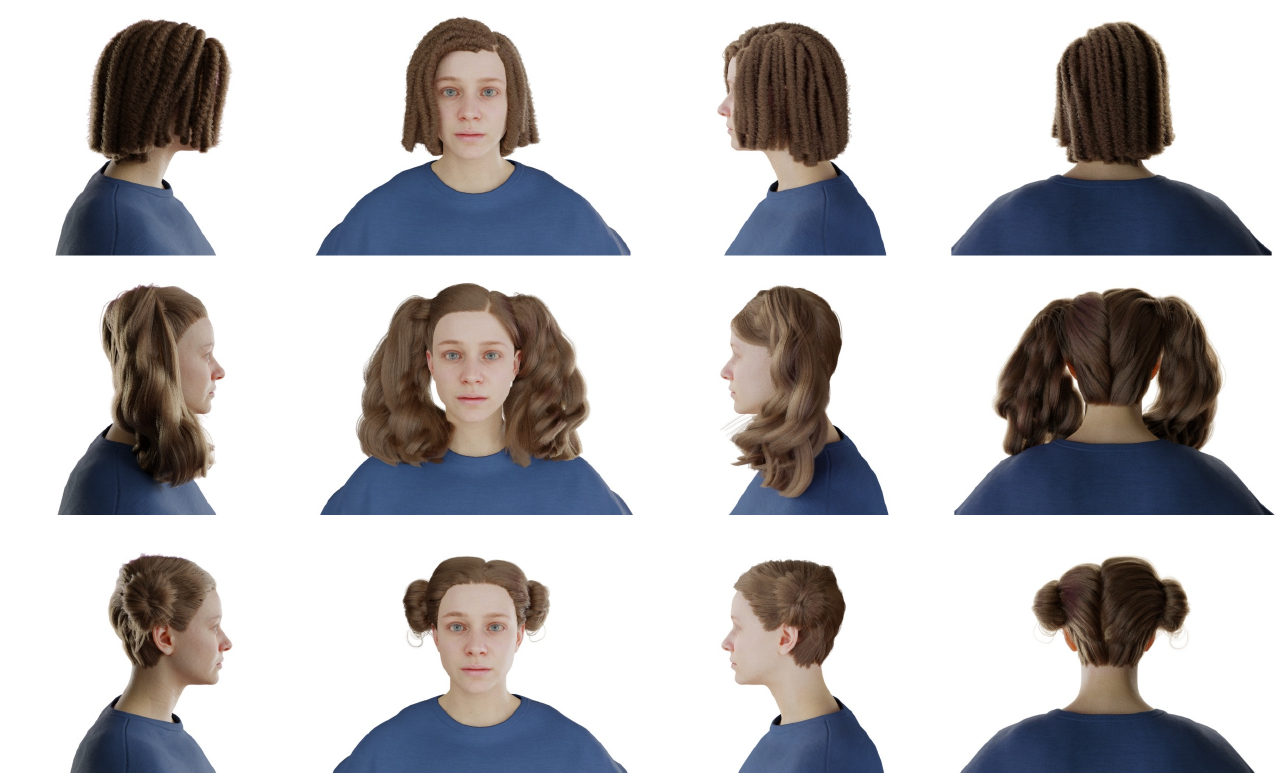}
    \caption{Examples of complex hairstyles in our database, including dreadlocks, double tails, and double buns.}
    \label{fig:database}
\end{figure}

%% file: fig/latex/retrieval.tex
\begin{figure}[ht]
    \centering
    \includegraphics[width=\columnwidth]{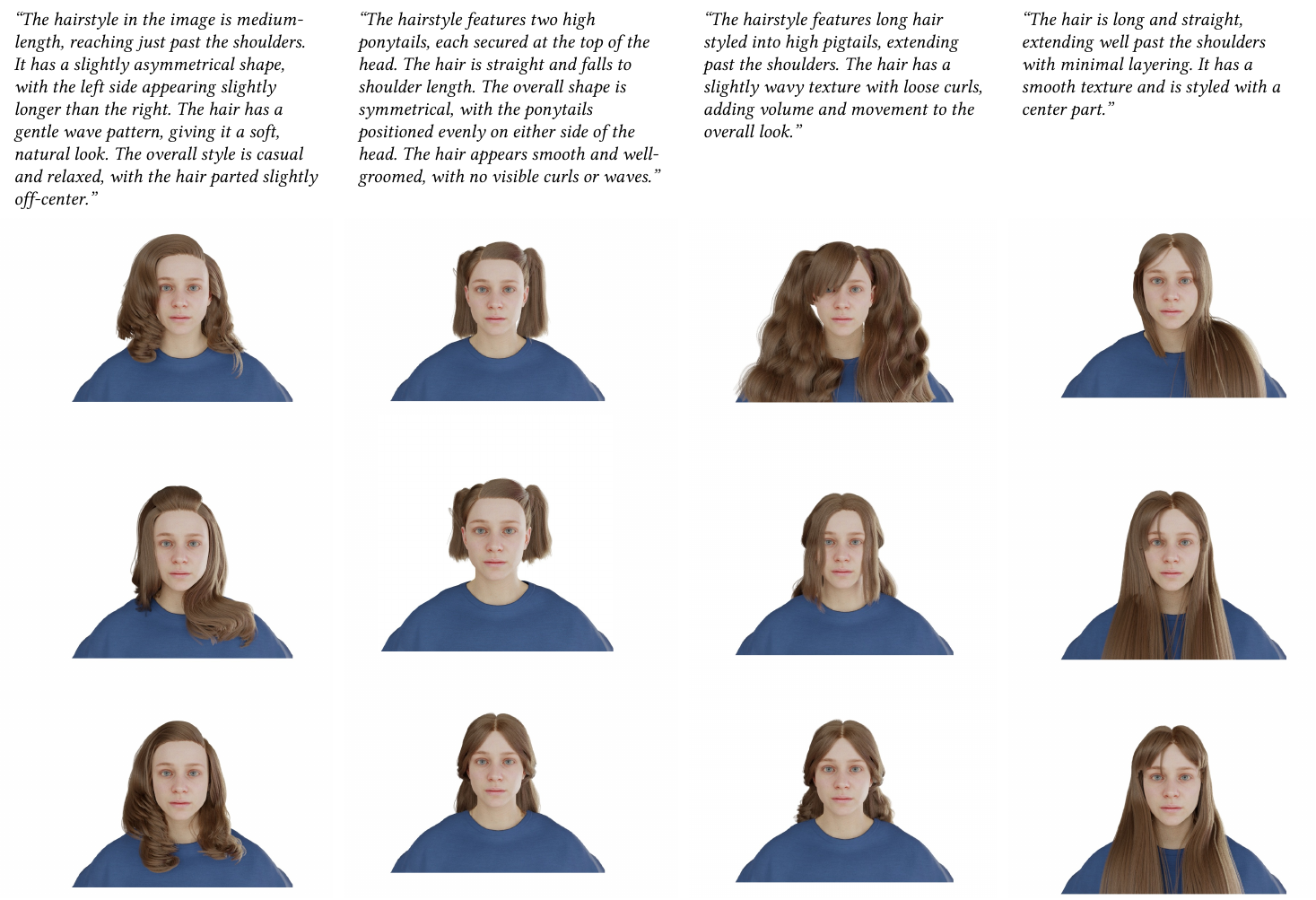}
    \caption{Examples of hairstyle candidates retrieved from our database based on user-provided textual descriptions. In each column, the top entry shows the input text prompt, followed by the three most relevant hairstyle candidates retrieved by our system, ordered from top to bottom by similarity.}
    \label{fig:retrieval}
\end{figure}

%% file: fig/latex/wind.tex
\begin{figure}[ht]
    \centering
    \includegraphics[width=\columnwidth]{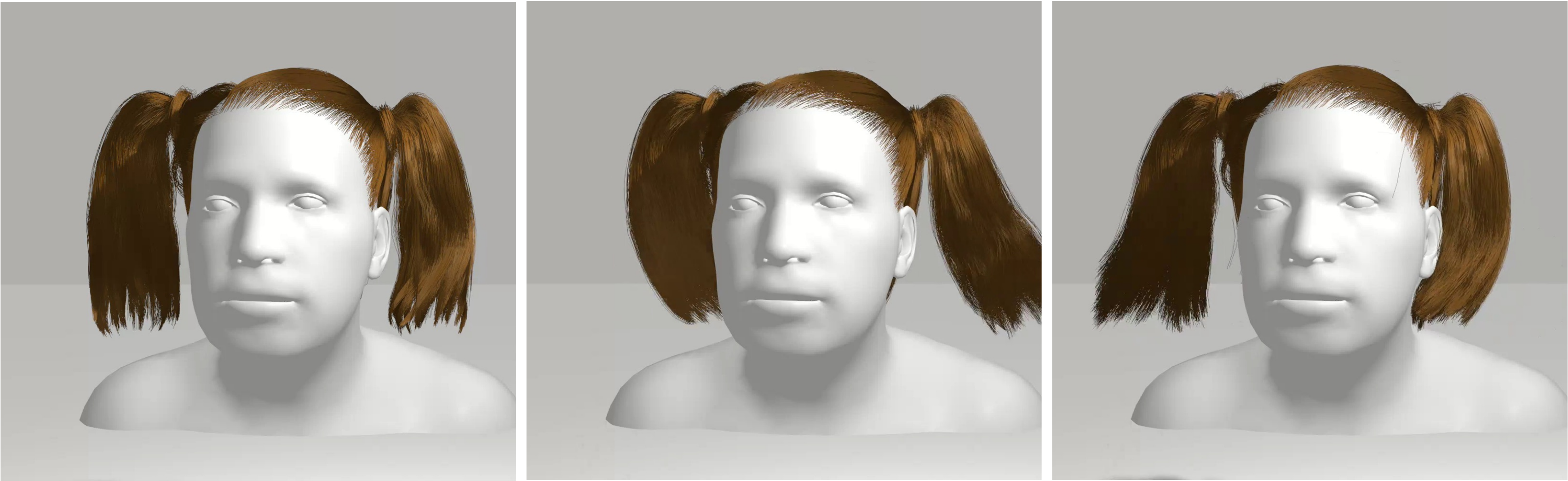}
    \caption{Real-time simulation of double tails under a wind field.}
    \label{fig:wind}
\end{figure}

%% file: algorithm/growth.tex
\begin{algorithm}[ht]
\KwIn{Growth steps $T$}
\Parameter{$p_\Gamma$, $p_\gamma$, $p_\Omega$, $p_h$, $p_{\text{freq}}$}
\KwOut{Strand vertices $\{\boldsymbol{v}_0, \boldsymbol{v}_1, \dots, \boldsymbol{v}_T\}$}

Initialize: Sample root position $\boldsymbol{v}_0$ within the triangle;
Sample perturbation vector $\boldsymbol{\delta} \sim \mathcal{U}(-1, 1)$\;
$\boldsymbol{p}^0_{\text{dir}} \gets$ initial direction from surface normal $+\,\boldsymbol{\delta}$\;

\For{$i \gets 1$ \KwTo $T$}{
    Compute gravity vector: $\boldsymbol{p}^i_{\text{grav}} \gets (0, -i p_\gamma, 0)$\;
    
    Update direction with gravity:
    \[
    \boldsymbol{p}^{{i}^{'}}_{\text{dir}} \gets \boldsymbol{p}^{i-1}_{\text{dir}} + \boldsymbol{p}^{i-1}_{\text{grav}}\,\max\left(p_{\Gamma}, 1-\|\boldsymbol{p}^{i-1}_{\text{dir}}\cdot (0,1,0)\|\right)
    \]
    
    Compute helix vector: 
    \[
    \boldsymbol{H}^i \gets \left(p_h \cos(i p_{\text{freq}}), 1, p_h \sin(i p_{\text{freq}})\right)
    \]
    
    Apply curl effect:
    \[
    \boldsymbol{p}^i_{\text{dir}} \gets \boldsymbol{p}^{i'}_{\text{dir}} + p_\Omega (\boldsymbol{p}^{i'}_{\text{dir}} - \boldsymbol{H}^i)
    \]
    
    Compute next vertex: $\boldsymbol{v}_i \gets \boldsymbol{v}_{i-1} + \boldsymbol{p}^i_{\text{dir}}$\;
}
\Return{$\{\boldsymbol{v}_0, \boldsymbol{v}_1, \dots, \boldsymbol{v}_T\}$}
\caption{\textbf{Procedural Strand Growth}}
\label{algo:growth}
\end{algorithm}

%% file: fig/latex/growth.tex
\begin{figure}[ht]
    \centering
    \includegraphics[width=\columnwidth]{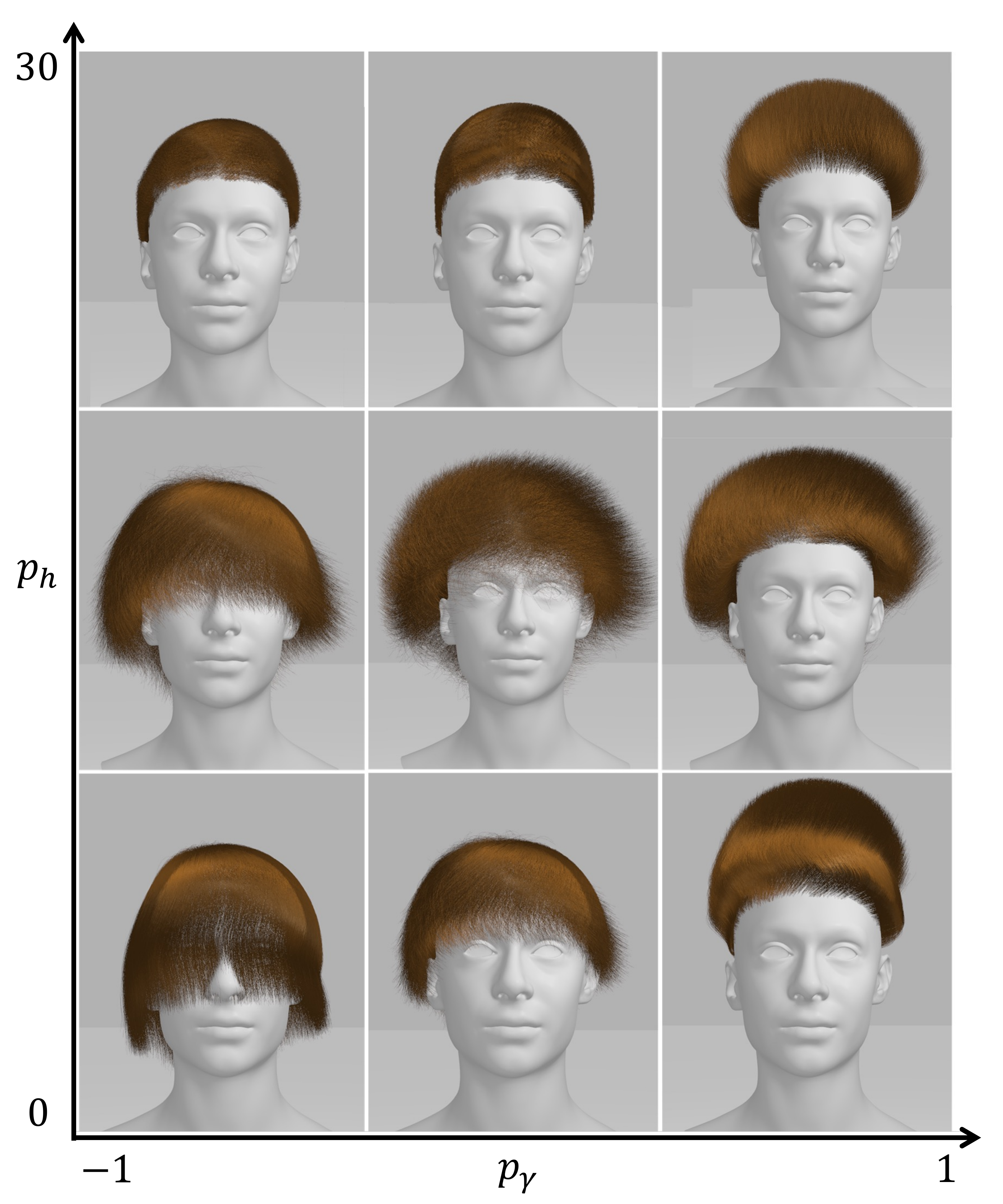}
    \caption{Parameter space exploration showing the impact of the helix radius $p_{h}$ and the gravity influence parameter $p_{\gamma}$ in our procedural hair growth algorithm.}
    \label{fig:growth}
\end{figure}

%% file: fig/latex/beard.tex
\begin{figure}[ht]
    \centering
    \includegraphics[width=\columnwidth]{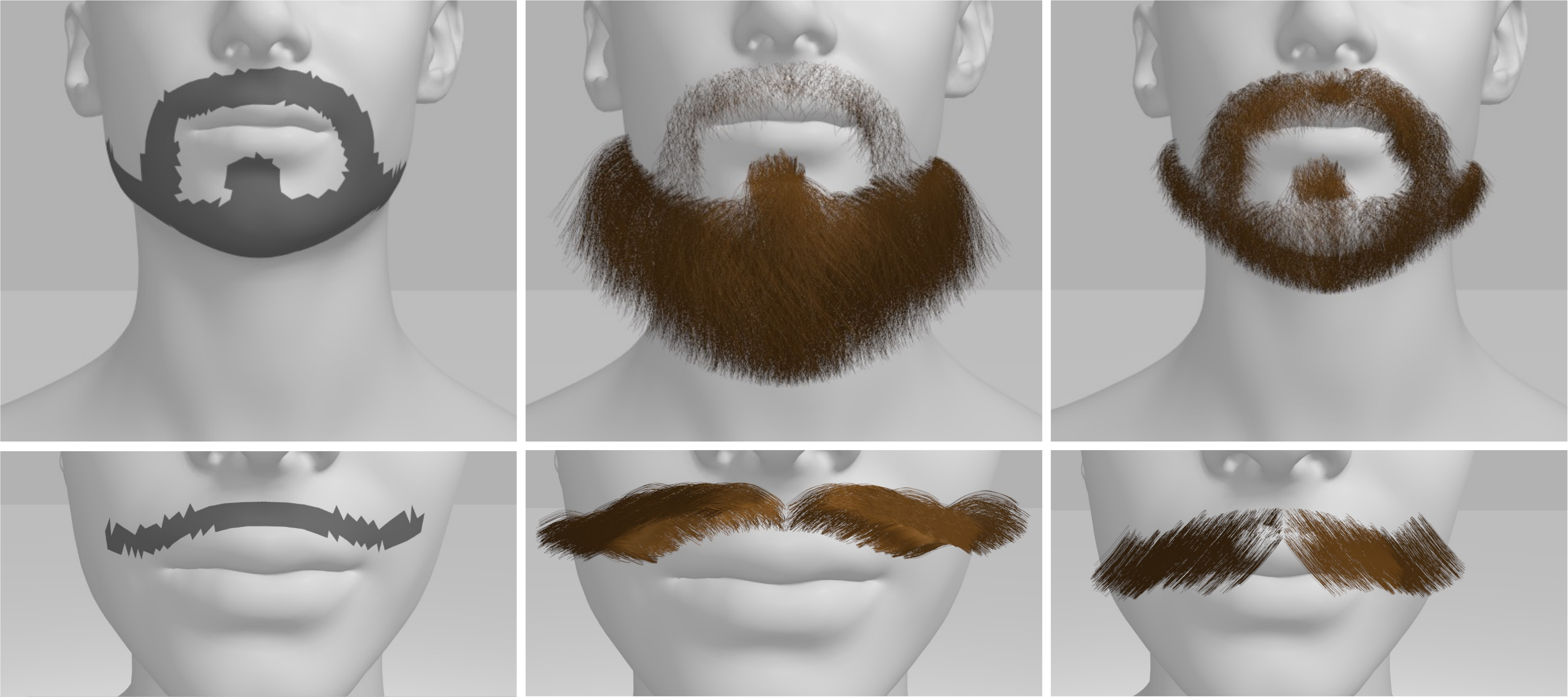}
    \caption{Procedural facial hair generation using our algorithm. Users specify growth regions by painting directly on the mesh surface (left), enabling intuitive creation of diverse facial hair styles (middle and right), including beards (top) and mustaches (bottom).}
    \label{fig:beard}
\end{figure}

%% file: fig/latex/trim.tex
\begin{figure}[ht]
    \centering
    \includegraphics[width=\columnwidth]{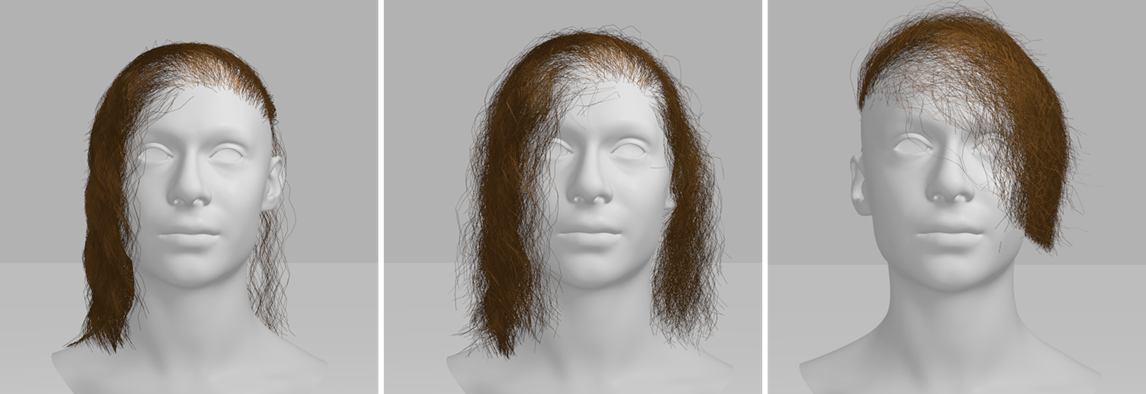}
    \caption{Simulation-driven grooming in our system. Beginning with an initial hair configuration (left), the user interacts with the hair by grabbing and displacing it to the left (middle), causing a dynamic, physically accurate response. In the final step (right), additional manipulation and trimming are performed, resulting in a shortened and restyled configuration. 
    }
    \label{fig:trim}
\end{figure}

%% file: fig/latex/controlnet.tex
\begin{figure}[ht]
    \centering
    \addtolength{\tabcolsep}{-4pt}
    \begin{tabular}{cccc}
     \includegraphics[width=0.242\columnwidth]{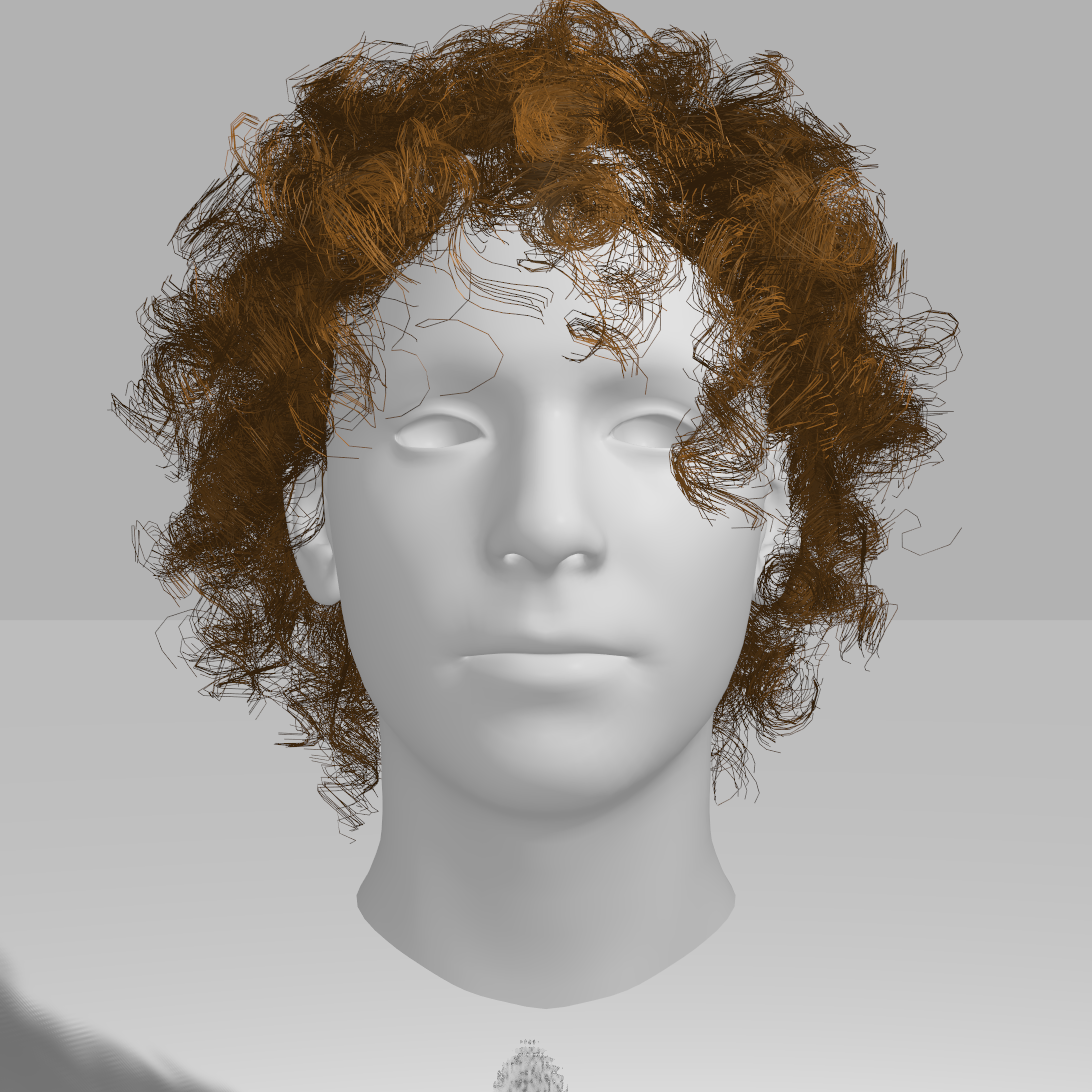} &
     \includegraphics[width=0.242\columnwidth]{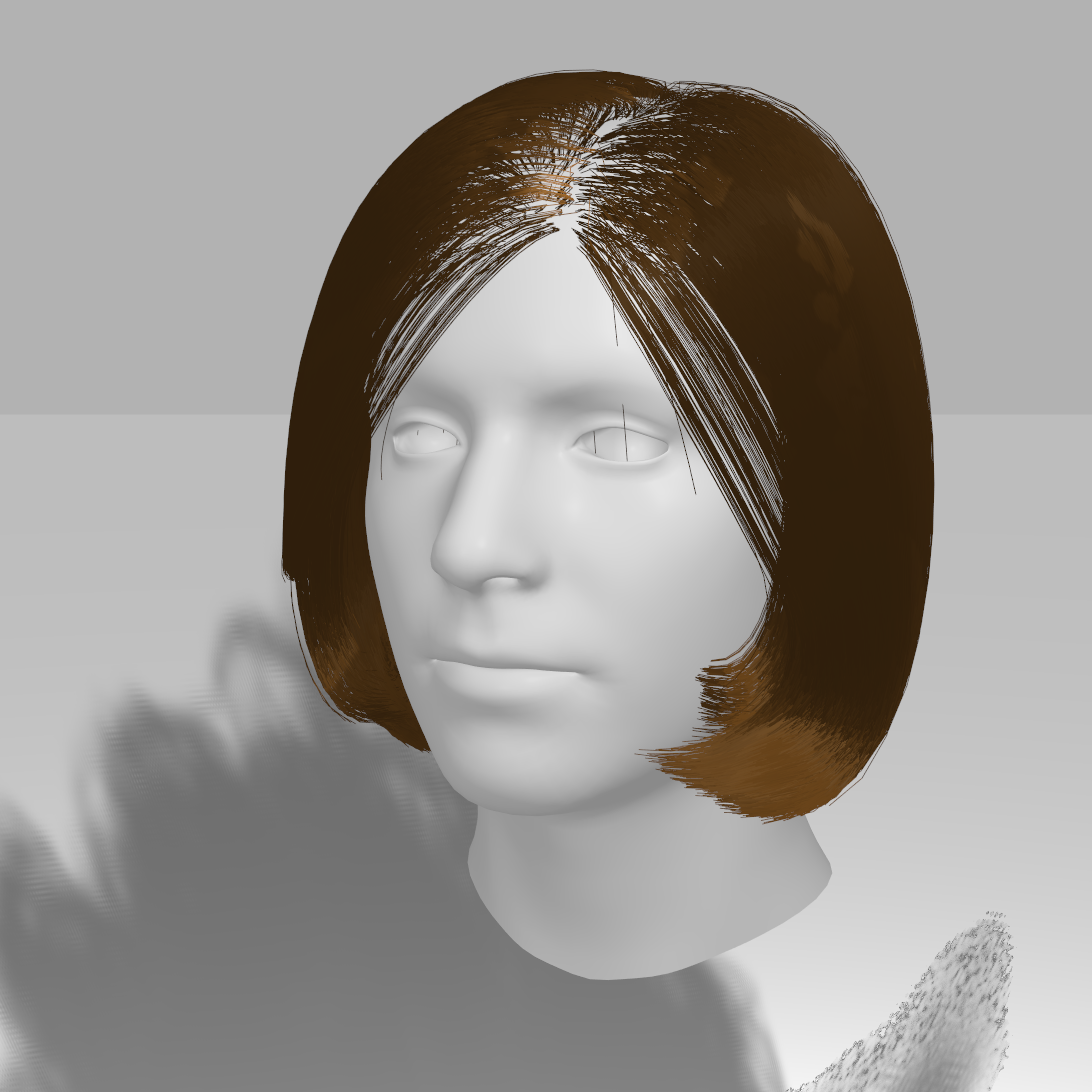} &
     \includegraphics[width=0.242\columnwidth]{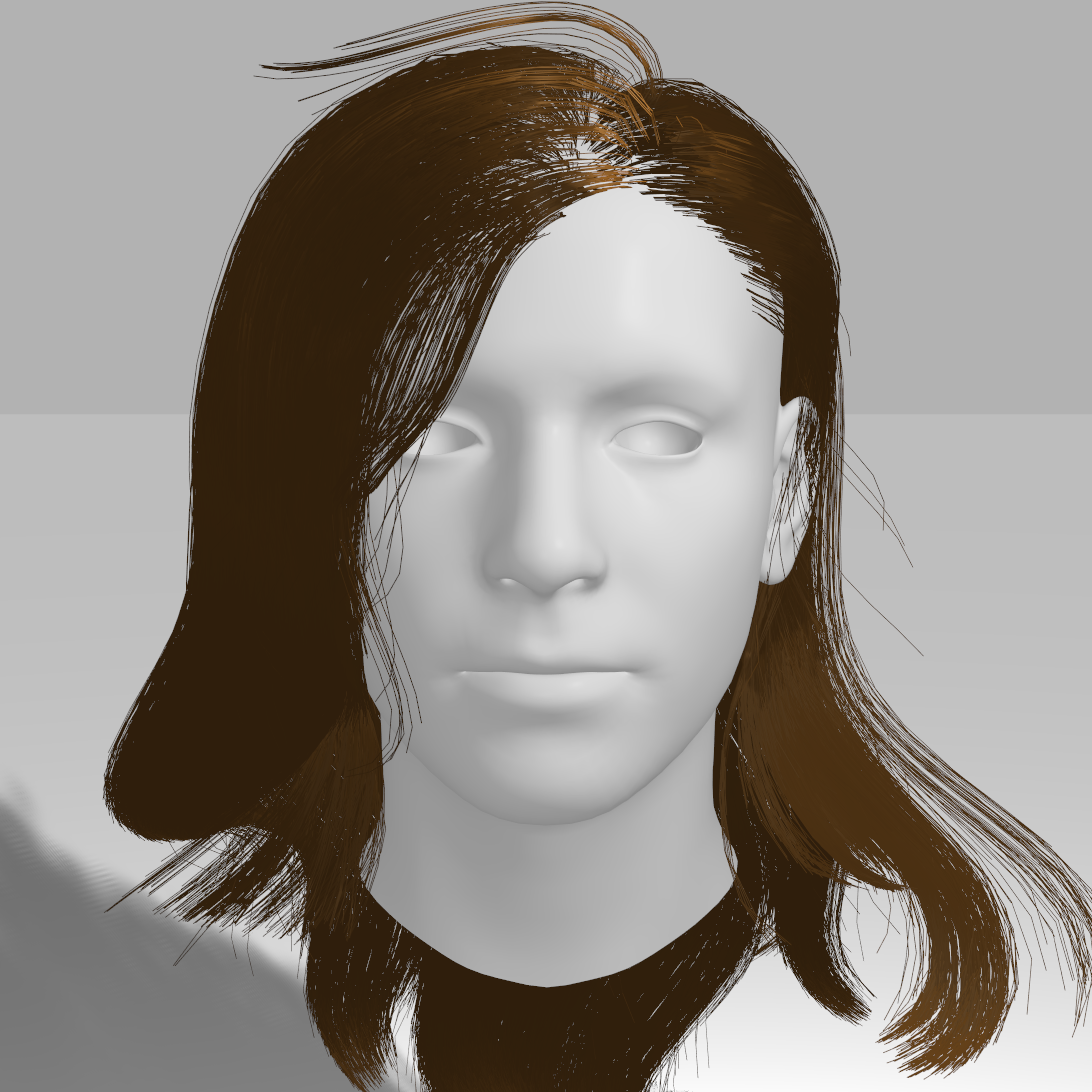} &
     \includegraphics[width=0.242\columnwidth]{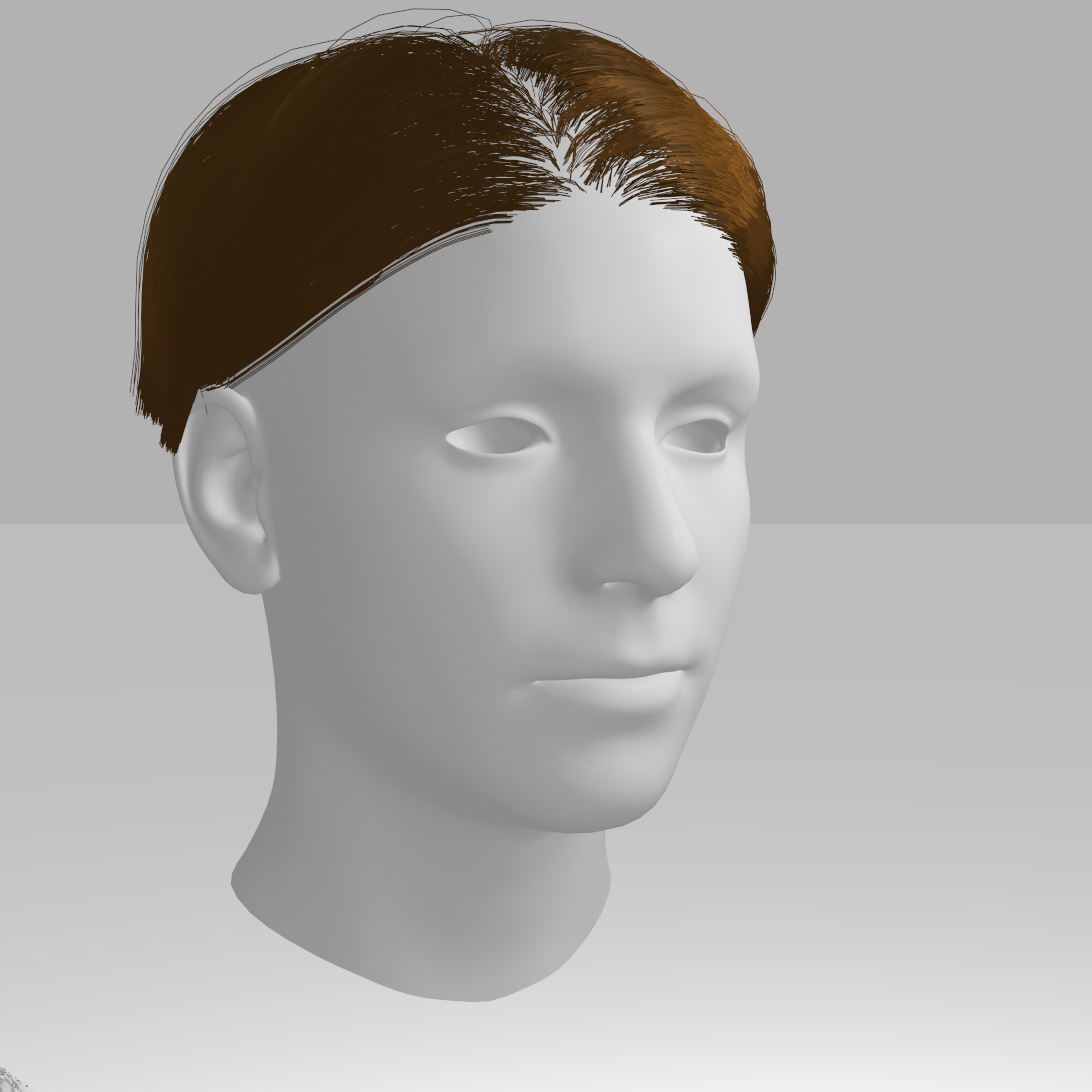} \\
     \frame{\includegraphics[width=0.242\columnwidth]{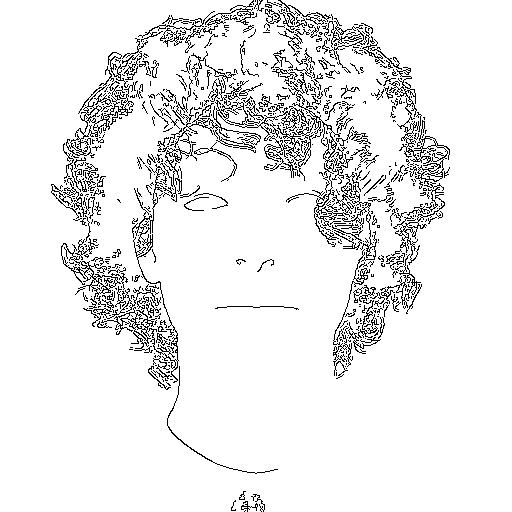}} &
     \frame{\includegraphics[width=0.242\columnwidth]{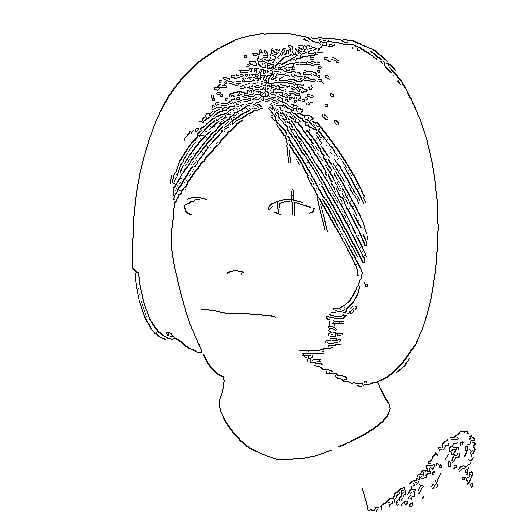}} &
     \frame{\includegraphics[width=0.242\columnwidth]{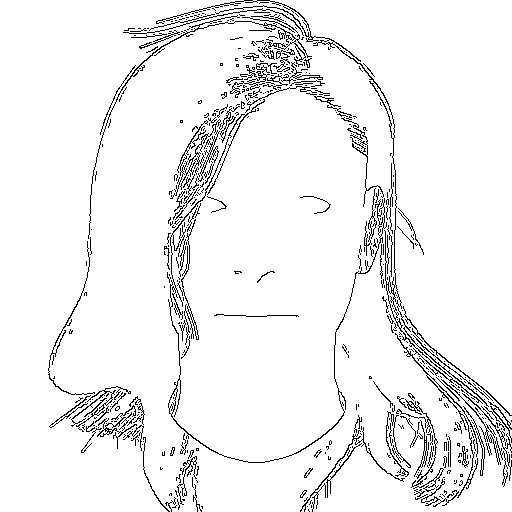}} &
     \frame{\includegraphics[width=0.242\columnwidth]{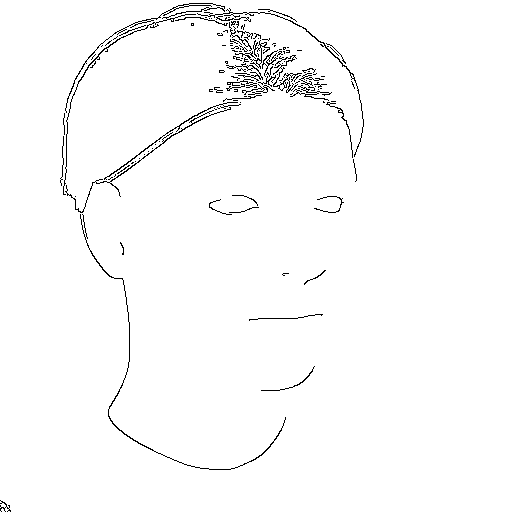}} \\
     \includegraphics[width=0.242\columnwidth]{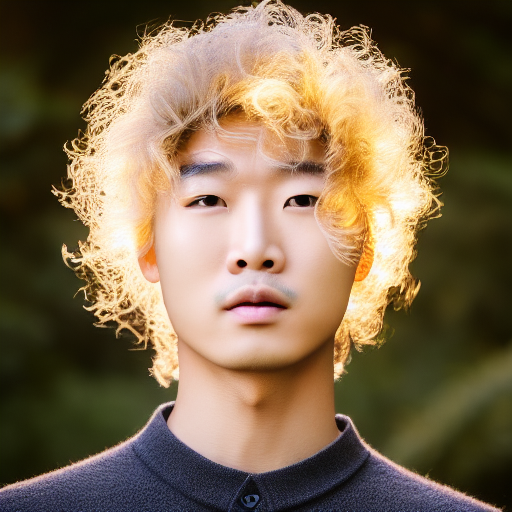} &
     \includegraphics[width=0.242\columnwidth]{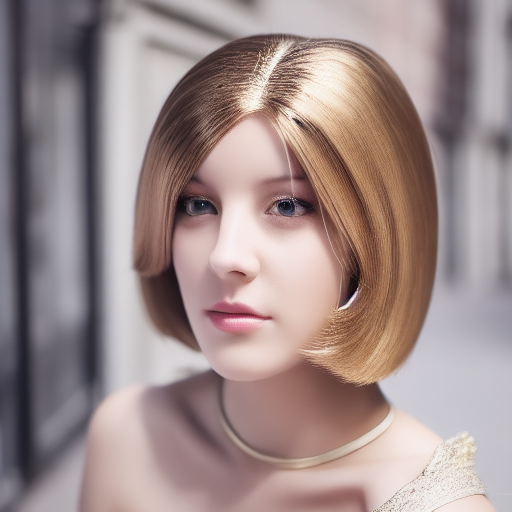} &
     \includegraphics[width=0.242\columnwidth]{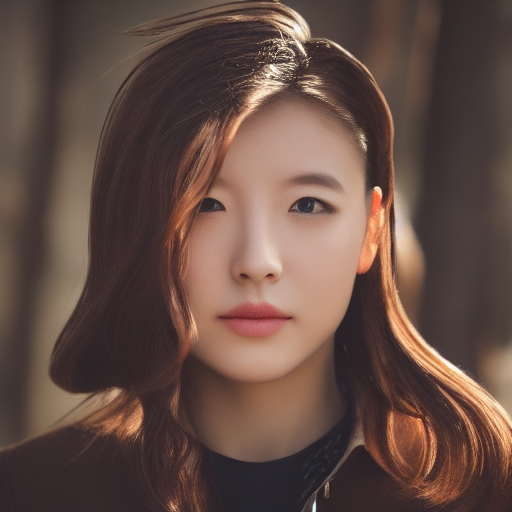} &
     \includegraphics[width=0.242\columnwidth]{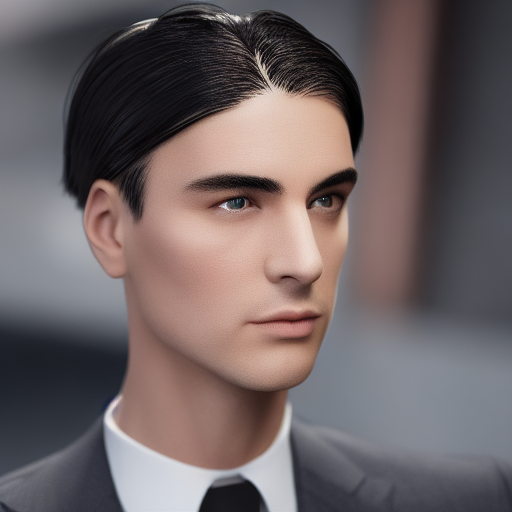} \\
    \end{tabular}
    \caption{Hair-conditioned image generation. From top to bottom, we show the cropped screenshot of the 3D viewport, the corresponding edge map extracted using the Canny filter, and the final photorealistic image generated by ControlNet. The following attributes are specified for each example in image generation: (1) \textit{``Asian male, blonde hair, frontal face, wear a sweater, in a park''}. (2) \textit{``Caucasian female, a short bob, slightly looking left, wear a dress''}. (3) \textit{``female, medium wavy, front face, wear a sweater''}. (4) \textit{``American male, short black hair, slightly looking right, wear a suit''}.}
    \label{fig:controlnet}
\end{figure}

%% file: sec/4_exp.tex
\section{Evaluation}

In this section, we evaluate how Digital Salon facilitates 3D hair modeling through two complementary studies: a quantitative assessment on the runtime performance of our system, and a user study collecting both quantitative metrics and qualitative feedback on usability and creative experience.

\subsection{Runtime Performance}
\label{sec:runtime}

To evaluate the runtime performance of Digital Salon, we benchmarked the latency of our key components (text-guided hair retrieval, real-time simulation, and hair-conditioned image generation) across various hairstyles.
All experiments were conducted on a consumer-grade machine equipped with an Intel\textsuperscript{\tiny\textregistered} Core\texttrademark\ i9-10850K CPU @ 3.60GHz, 64GB RAM, and an NVIDIA RTX 3090 GPU with 24GB VRAM.

\input{table/performance}

We report the average latency over five trials for each hairstyle configuration in~\cref{tab:performance}. For a typical hairstyle consisting of approximately $10,000$ strands, our system takes around 3 seconds to retrieve the corresponding model from the database. The simulation component maintains real-time performance, with all configurations running under $20$ ms per frame -- equivalent to over $50$ FPS on our test hardware. The most time-consuming component is image generation, which takes approximately $22$ seconds to generate an image of resolution $512 \times 512$.
While this may appear slow, traditional physics-based rendering pipelines require significantly more effort and time. For example, rendering a similar visual result in Blender takes over $40$ seconds on the same hardware, excluding manual modeling and setup time for additional elements like clothing and accessories.
Besides, our current implementation is not fully optimized. At present, communication between the interface and backend relies on Python/C API, which reloads ControlNet with every image generation. This could be improved by using Python bindings with C++ and initializing ControlNet only once at startup. Additionally, implementing a queuing mechanism would allow image generation to happen asynchronously in the background. We leave these enhancements as future engineering work, as the primary goal of Digital Salon is to demonstrate the conceptual feasibility rather than to address low-level system optimization issues.

In terms of resource usage, the peak memory consumption during typical usage reaches $24.96$GB of system RAM and $9.24$GB of VRAM, primarily during the image generation phase involving ControlNet. Based on our study, this memory consumption is within the capabilities of many modern consumer-grade desktops and high-end laptops.

In summary, these evaluations prove the real-time performance of our system on consumer-grade hardware, allowing users to efficiently generate, simulate, and render 3D hairstyles with minimal latency.

\subsection{User Study}

\subsubsection{Participants}

We recruited seven participants (P1 -- P7), including five males and two females, aged between $26$ and $40$. Among them, two participants (P3 and P4) had prior experience in 3D hair modeling, three participants (P2, P5, and P7) had general experience in 3D content creation, and the remaining two participants (P1 and P6) were novices in these areas. All seven participants were invited to an in-person user study, where they interacted directly with our system following the procedure outlined below. We collected both qualitative feedback and quantitative ratings on specific aspects of the system.
Additionally, we conducted a remote user study with a professional stylist (P8). Due to the remote format, we operated the system on P8's behalf. Although we did not collect quantitative ratings from P8 due to the lack of hands-on interactions, we gathered qualitative feedback on the system’s usability in real-world scenarios.

\subsubsection{Procedure}

We conducted the study with each participant individually. Each study consisted of the following three sessions.

\textbf{Introduction and training session (10 minutes).} Following a brief background check, participants were introduced to the project and given a tutorial on the system. This included pre-recorded instructional videos and a reference menu outlining all supported operations. Participants were asked to review the tutorial materials and encouraged to replicate the demonstrated tasks to familiarize themselves with the system. These tasks covered all major components: text-guided hair retrieval, real-time hair simulation, interactive grooming, and hair-conditioned image generation. Participants were free to ask questions and explore the system until they felt confident in using the system.

\textbf{Free hairstyle exploration (20 minutes).} Participants were invited to freely create several hairstyles based on their own imagination, without any constraints or specific instructions. This open-ended task encouraged creativity and allowed participants to explore their personal styling preferences. During this phase, they were also asked to engage with the full functionality of our system, including hair generation, simulation, and rendering, to help identify potential usability issues and evaluate the system’s overall performance.

\textbf{Post-study survey and interview (10 minutes).} After the exploration session, participants were asked to share their overall impressions of the system, including any strengths and weaknesses they observed. They then completed a seven-point Likert questionnaire based on their experience, rating the system across six criteria: learning curve, intuitiveness, latency, precision, enjoyment, and overall preference. Finally, participants were asked to suggest potential application scenarios where they could envision using Digital Salon in the future.

\subsection{Results and Findings}

\cref{tab:user-results} summarizes the hairstyles created by each participant along with the average time spent during the study, and~\cref{fig:rating} presents user evaluations across several quantitative metrics. Below, we discuss detailed findings related to specific aspects of our system based on participant feedback.
\input{table/user_results}

\input{fig/latex/rating}

\subsubsection{Intuitiveness and Efficiency}

Regarding the \textbf{learning curve} and \textbf{intuitiveness} of our system, most participants reported that the system was easy to learn and intuitive to use, especially with the aid of the provided instructional videos. Only $1$ out of $7$ participants rated the learning curve as $4$, and $2$ out of $7$ rated the intuitiveness as $4$. P1, a novice user, commented, \textit{``It is pretty intuitive to use. You have text input, and you have a bunch of controls for the hair.''} P4 appreciated the 3D nature of the system, stating, \textit{``The operations are pretty straightforward, and being a 3D system, I can view the hairstyle from arbitrary perspectives.''} P4 also emphasized the value of text input, noting, \textit{``Text provides a good baseline for users, especially for those who have limited background in 3D hair modeling, as they now do not need to draw the hair from scratch.''}
However, P6, another novice user, expressed some concerns about the system's intuitiveness, saying, \textit{``The left parameter window exposes lots of detailed parameters for high-end users, which is actually a bit distracting for me.''} Fortunately, this issue is easy to address, since all floating windows in our system can be minimized. We thus left the detailed parameter panel to experienced users, while letting novices to focus solely on the interaction panel.
In summary, user feedback on learning curve and intuitiveness shows a positive trend and supports our claim that Digital Salon is an intuitive hair authoring system.

For the \textbf{latency} of our system, most participants reported that the text-guided hair retrieval and real-time hair simulation modules were responsive and ran smoothly without noticeable lag. However, some concerns were raised about the hair-conditioned image generation component, which takes approximately $22$ seconds per image. This delay was perceived as slightly slow, particularly by participants without prior experience in physics-based hair rendering (P2 and P6). As discussed in~\cref{sec:runtime}, a 20-second delay is generally acceptable given the complexity of this task, and we suggested certain engineering optimizations that could be applied to improve image generation efficiency. Overall, participants still acknowledged the system’s strong performance and responsiveness across most stages of the workflow.
\cref{tab:user-results} lists the average time each participant spent designing a hairstyle, including the stages of generation, simulation, grooming, and final rendering. Although the time varied according to the technical backgrounds of the participants, all users were able to complete the modeling process in less than $10$ minutes. This is a significant improvement over traditional hair prototyping workflows that require several days for realistic and visually compelling hairstyles, and is comparable to the results reported by Xing et al.~\cite{xing2019hairbrush}, whose system focused solely on hair generation in VR.
Taken together, the user feedback and the runtime performance reported in~\cref{tab:performance} support our claim that Digital Salon is an efficient and interactive system for hair authoring.

\subsubsection{Precision}

To evaluate the precision of our system, we asked participants to rate two aspects separately: (1) the \textbf{precision between the retrieved 3D hairstyles and the input text prompts}, and (2) \textbf{the precision between the generated images and the textual descriptions provided}. Regarding hair retrieval, most participants agreed that the 3D hairstyles retrieved from our curated database matched their imagination of the described prompt. As P6 noted, \textit{``Many parts of my input in the prompts have been reflected in what I see in the output.''}
However, some edge cases remained difficult to cover, particularly when participants experimented with weird or unconventional prompts (P2 and P5). In such cases, our interactive grooming tool proved valuable, allowing participants to refine mismatched details in the retrieved hairstyles. For example, P4 found that the initially retrieved hairstyle was basically matched but with some discrepancy left in certain details. After refining it using our grooming tool, the final result matched the expectation exactly.
In summary, our text-guided hair retrieval module provides a level of precision sufficient for most participants' needs, while the grooming tool serves as an effective complement to address fine-grained or missing details.

For image generation, participants reported higher perceived precision compared to hair retrieval, highlighting the power and flexibility of ControlNet in adapting to various textual prompts and structural conditions. Both P1 and P3 appreciated the ability to include clothing during image generation. As P1 remarked, \textit{``Being able to add the clothing part is nice. It gives a more complete picture of what it would look like on the person instead of just a floating head.''}
In summary, this feedback confirms that our system can produce photorealistic previews of 3D hairstyles within a reasonable time range, with results that align well with both the hair structure and textual descriptions. Additionally, our system is designed to be agnostic to the underlying image generation backbone, meaning that it can be easily adapted to leverage more advanced models such as Adobe Firefly~\cite{firefly}, enabling further improvements in alignment accuracy and visual fidelity.

\subsubsection{Experience and Implications}

\textbf{All participants agreed that our system was enjoyable to use}. P1 noted, \textit{``The granularity of the hair you can see is good. You can see enough hair to kind of imagine what the photo would look like, and you still have control over individual strands.''} P3 stated, \textit{``I like the physics. Adding physics onto every step made it really cool. Also, I like the brush, making it able to add the beard everywhere.''} P6, although expressed some concerns about the system’s intuitiveness, acknowledged the engaging experience offered through the chat-based interaction, remarking, \textit{``With the medium of chatting, it is easy and fun to control from the user's perspective, and playing with hair on human characters is enjoyable.''}
Together, these responses suggest that Digital Salon provides a fun, engaging, and creative experience for users with diverse backgrounds and levels of technical expertise. In~\cref{fig:results} we show some example designs by participants in the user study.
\input{fig/latex/results}

Regarding preferences between Digital Salon and professional 3D hair modeling software such as Blender, participants expressed a range of perspectives. Some highlighted the advantages of our system; for example, P2 noted, \textit{``Some operations in this system are hard to do in Blender.''} Others emphasized the potential for complementary usage rather than replacement. As P3 remarked, \textit{``If I could generate something in Blender and have the physics here, I think that would be really good. For stuff like hair nodes, I think Blender is still more powerful, because right now I can only do operations like chatting and cutting here.''} Similarly, P5 commented, \textit{``If you want to focus on playing with hair, it might be enough. But for more complex operations, software like Blender may be better suited.''}
However, it is worth noting that \textbf{Digital Salon is not intended to replace professional 3D software, but rather to augment existing workflows by lowering the barrier to entry and enabling rapid prototyping}. Based on participants' feedback, Digital Salon achieves this goal effectively.

The potential use case of Digital Salon was further supported by P8, a professional stylist, who noted that the system offers a more effective medium than traditional magazine reference images. By incorporating 3D hairstyles, realistic hair simulation, and real-time interactivity, Digital Salon can provide clients with a more accurate and personalized preview of their desired styles. As suggested by P8, the system could be deployed in salons to assist clients in designing and selecting hairstyles they wish to try. Clients would benefit from being able to quickly test their ideas with previews that reflect the physical properties of their own hair. Salon staff could provide high-level guidance, with final checks and approval made by a professional stylist. According to P8, this workflow would enhance efficiency and minimize the confusion and miscommunication that often occur in conventional stylist-client interactions.

%% file: table/performance.tex
\begin{table*}[ht]
\centering
\caption{Runtime performance of major components in Digital Salon, evaluated across three representative hair types. Values are reported as mean $\pm$ standard deviation over five trials.}
\label{tab:performance}
\addtolength{\tabcolsep}{2pt}
\begin{tabular}{@{}cccccc@{}}
\toprule
\textbf{Hairstyle} & \textbf{Text Prompt}     & \textbf{\# Strands} & \textbf{Retrieval Time (s)} & \textbf{Simulation Time (ms/frame)} & \textbf{Image Generation Time (s)} \\ \midrule
Short              & \textit{``short bob''}   & $8412$              & $2.89\pm0.05$               & $12.61\pm0.15$                     & $21.74\pm0.29$                     \\
Medium             & \textit{``medium wavy''} & $9833$              & $3.00\pm0.22$               & $16.02\pm0.14$                     & $21.75\pm0.08$                     \\
Long               & \textit{``long curly''}  & $9132$              & $2.89\pm0.10$               & $19.04\pm0.05$                      & $21.90\pm0.39$                     \\ \bottomrule
\end{tabular}%
\end{table*}

%% file: table/user_results.tex
\begin{table*}[ht]
\caption{A summary of time spent and the design concept developed by each participant in the in-person study.}
\label{tab:user-results}
\addtolength{\tabcolsep}{3pt}
\begin{tabular}{@{}ccl@{}}
\toprule
\# & \textbf{Average Modeling Time} & \textbf{Design Concept} \\ \midrule
P1 & $7$ min $22$ s & Long layered hair with loose waves and short bangs. \\
P2 & $3$ min $45$ s & A hairstyle of a person who has a psychiatric illness and has cut their own hair in a very weird way. \\
P3 & $7$ min $21$ s & A mohawk hairstyle. \\
P4 & $5$ min $45$ s & A long wavy hair, with $3$cm radius curls, longer than shoulder length. \\
P5 & $5$ min $25$ s & A monkey-like hairstyle being blown away by wind. \\
P6 & $5$ min $35$ s & A short-sided hairstyle with a clipper cut. \\
P7 & $9$ min $50$ s & A straight and thick female hairstyle. \\ \bottomrule
\end{tabular}%
\end{table*}

%% file: fig/latex/rating.tex
\begin{figure}[ht]
    \centering
    \includegraphics[width=\columnwidth]{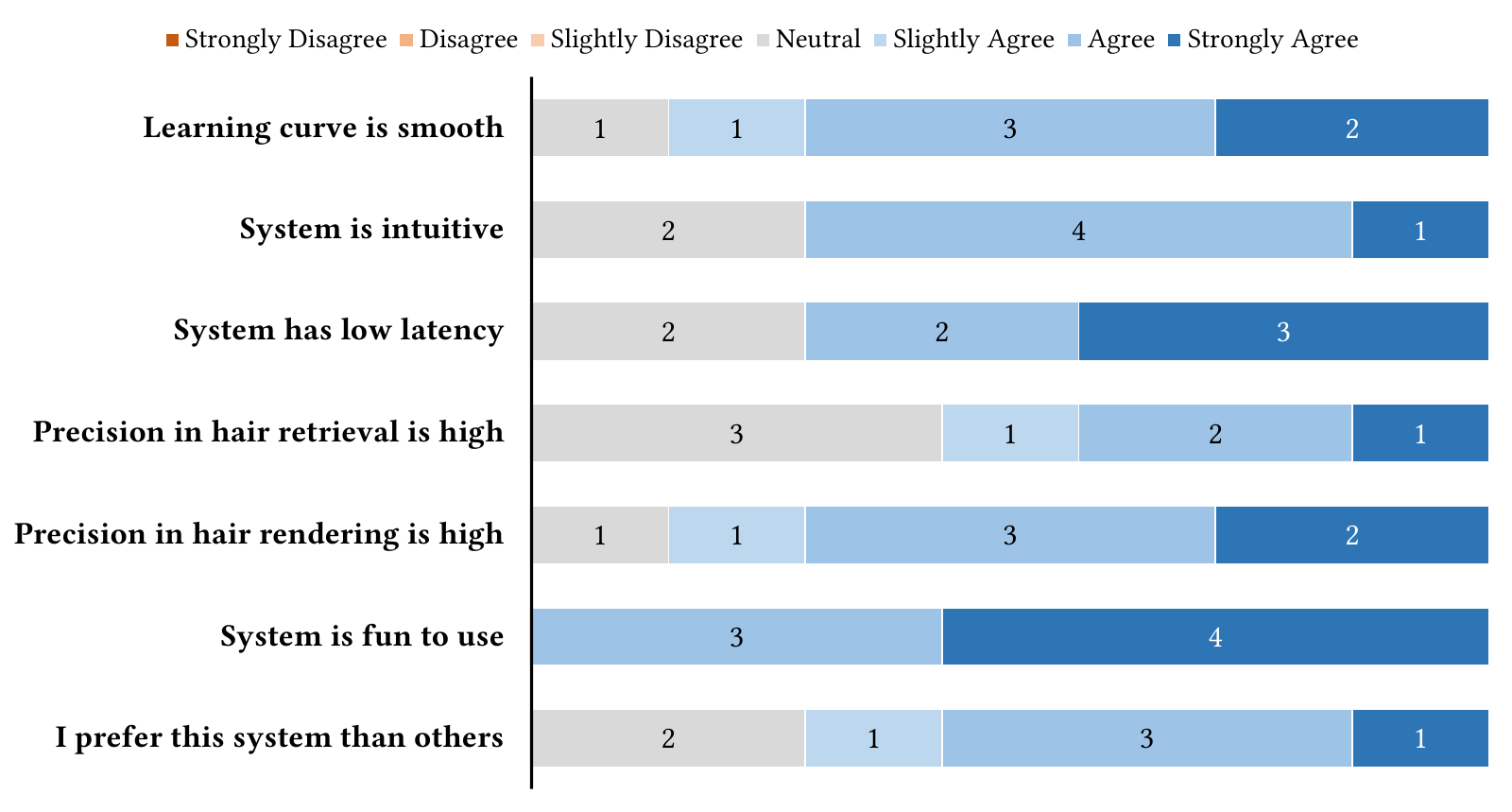}
    \caption{Participant ratings across six evaluation metrics: learning curve, intuitiveness, latency, precision, enjoyment, and overall preference.}
    \label{fig:rating}
\end{figure}

%% file: fig/latex/results.tex
\begin{figure*}[ht]
    \centering
    \includegraphics[width=\linewidth]{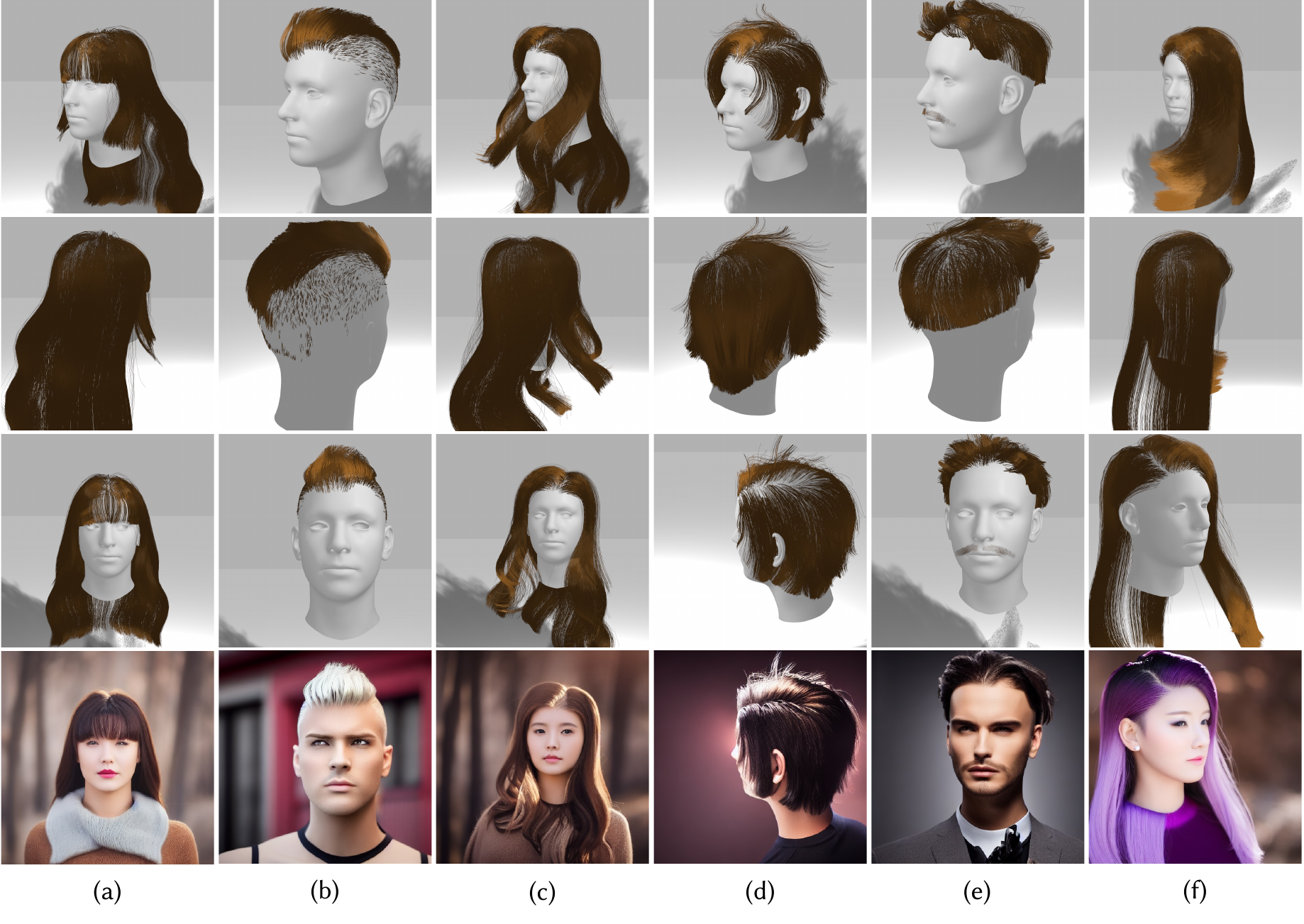}
    \caption{Example designs by participants in the user study. (a) P1 designed a long hairstyle with loose waves and short bangs. (b) P3 created a mohawk-style look that was not present in the database and used the grooming tools to manually trim the retrieved result. (c) P4 designed a long, wavy hairstyle with curls. (d) P5 created a playful, monkey-inspired hairstyle; even from a back view, the rendered image maintained high visual quality. (e) P6 designed a short-sided hairstyle and added painted mustaches. (f) P7 created a long, straight hairstyle and rendered it in a rare but striking purple color.}
    \label{fig:results}
\end{figure*}

%% file: sec/5_conclusion.tex
\section{Limitations and Future Work}

Although our study validated the intuitiveness and effectiveness of Digital Salon in facilitating 3D hair modeling, there are limitations and many areas left for future work.

First, our post-study interviews revealed a strong desire for a more \textbf{personalized experience} in Digital Salon. Both P2 and P3 expressed interest in seeing their own face in the 3D viewport and in the generated images, rather than the default mesh template or randomly synthesized faces currently shown in the images. A similar concern was raised by P8, who emphasized that in real-world scenarios, face shapes play a crucial role in determining the suitability of a hairstyle. For example, clients with longer face shapes may not suit short hairstyles, which can appear visually unbalanced or less flattering.
To address this, future iterations of Digital Salon could incorporate face fitting algorithms such as DELTA~\cite{feng2023learning}, which enable the reconstruction of personalized 3D head meshes from portrait images uploaded by users. For rendering more personalized visual results, techniques like DreamBooth~\cite{ruiz2023dreambooth} could be explored, allowing a general diffusion model to be fine-tuned with just a few images of a specific individual. However, the computational cost of fine-tuning diffusion models remains a challenge for maintaining real-time performance. Addressing this limitation may require further research into more efficient training and inference pipelines.

While Digital Salon currently supports a range of hair grooming tools and algorithms, many \textbf{specialized tools commonly used in real-world salons} do not yet have digital counterparts in our system. For example, combs are often used to guide hair flow in specific directions, curling irons or rollers are used to increase curliness or introduce volume, and hair wax helps cluster strands to create and hold particular shapes. These tools are essential for precise hairstyling and artistic expression, yet they remain unsupported in the current version of Digital Salon. Incorporating digital versions of such tools represents a promising direction for future work and would contribute to making the virtual hair authoring experience more immersive and expressive, closely mirroring real-world salon practices.

In real-world salon visits, clients often bring reference images of desired hairstyles. However, Digital Salon currently does not support \textbf{image-guided hair retrieval}, relying instead on textual descriptions as input. The primary challenge lies in the variability of in-the-wild reference images, which often include differences in head pose, illumination, hair color, and occlusions. These factors make it difficult to extract robust image embeddings that focus solely on hair structure features, while factoring out all irrelevant visual information. As a result, we opted to use text as the primary guidance modality in our system.
Enabling image-guided hair retrieval thus remains an open and promising direction for future work, as it would enhance the applicability of Digital Salon in real-world scenarios and better align with common hairstyling practices.

Last but not least, \textbf{expanding the accessibility of Digital Salon to mobile platforms} such as smartphones and tablets would significantly broaden its usability in both consumer and salon settings. However, our current implementation relies heavily on CUDA to achieve real-time performance, which restricts its deployment to environments equipped with dedicated GPUs. Adapting the system to operate efficiently on devices with limited computational resources remains a critical challenge, which is also an interesting direction for future work.

\section{Conclusion}

In conclusion, Digital Salon presents a comprehensive hair authoring system that seamlessly integrates text-to-hair generation, physics-based simulation, and AI-driven rendering. The system is designed to achieve real-time performance on consumer-grade hardware, enabling users to rapidly prototype 3D hairstyles with interactive editing and dynamic simulation. Our user study demonstrates that Digital Salon effectively enhances creative exploration through intuitive interaction and real-time feedback. Furthermore, it shows promise for use in real salon environments, offering a more expressive and informative alternative to traditional reference images.
Looking ahead, future work will focus on improving personalization features, incorporating a broader range of virtual grooming tools, and supporting image-based hairstyle retrieval and mobile platforms. These extensions aim to make Digital Salon an even more powerful tool for both novice users and professional stylists, further bridging the gap between digital prototyping and real-world hairstyling.